\documentclass[aip,amsmath,amssymb,reprint,floatfix]{revtex4-2}
\usepackage{graphicx}%
\usepackage{dcolumn}%
\usepackage{bm}%
\usepackage{multirow}

\usepackage[utf8]{inputenc}
\usepackage[T1]{fontenc}
\usepackage{mathptmx}
\usepackage[table]{xcolor}
\usepackage{hyperref}
\hypersetup{
	colorlinks   = true, 
	urlcolor     = blue,
	linkcolor    = blue, 
	citecolor   = blue 
}
\newcommand{\etal}{\textit{et al.}}

\begin{document}

\title{General embedded cluster protocol for accurate modeling of oxygen vacancies in metal-oxides}

\author{Benjamin X. Shi}
\affiliation{Yusuf Hamied Department of Chemistry, University of Cambridge, Lensfield Road, Cambridge CB2 1EW, United Kingdom}%

\author{Venkat Kapil}
\affiliation{Yusuf Hamied Department of Chemistry, University of Cambridge, Lensfield Road, Cambridge CB2 1EW, United Kingdom}%
\affiliation{Churchill College, University of Cambridge, Storey's Way, Cambridge CB3 0DS}%

\author{Andrea Zen}
\affiliation{Dipartimento di Fisica Ettore Pancini, Universit\`{a} di Napoli Federico II, Monte S. Angelo, I-80126 Napoli, Italy}
\affiliation{Department of Earth Sciences, University College London, Gower Street, London WC1E 6BT, United Kingdom}

\author{Ji Chen}
\affiliation{School of Physics, Peking University, Beijing, 100871, China}

\author{Ali Alavi}
\affiliation{Max Planck Institute for Solid State Research, Heisenbergstra{\ss}e 1, 70569 Stuttgart, Germany}
\affiliation{Yusuf Hamied Department of Chemistry, University of Cambridge, Lensfield Road, Cambridge CB2 1EW, United Kingdom}%

\author{Angelos Michaelides}
\affiliation{Yusuf Hamied Department of Chemistry, University of Cambridge, Lensfield Road, Cambridge CB2 1EW, United Kingdom}%
\affiliation{Department of Physics and Astronomy, University College London, Gower Street, London, WC1E 6BT, United Kingdom}
\affiliation{Thomas Young Centre and London Centre for Nanotechnology, 17-19 Gordon Street, London WC1H 0AH, United Kingdom}

\date{\today}

\begin{abstract}
The O vacancy (Ov) formation energy, $E_\textrm{Ov}$, is an important property of a metal-oxide, governing its performance in applications such as fuel cells or heterogeneous catalysis. These defects are routinely studied with density functional theory (DFT). However, it is well-recognized that standard DFT formulations (e.g.\ the generalized gradient approximation) are insufficient for modeling the Ov, requiring higher levels of theory. The embedded cluster method offers a promising approach to compute $E_\textrm{Ov}$ accurately, giving access to all electronic structure methods. Central to this approach is the construction of quantum(-mechanically treated) clusters placed within suitable embedding environments. Unfortunately, current approaches to constructing the quantum clusters either require large system sizes, preventing application of high-level methods, or require significant manual input, preventing investigations of multiple systems simultaneously. In this work, we present a systematic and general quantum cluster design protocol that can determine small converged quantum clusters for studying the Ov in metal-oxides with accurate methods such as local coupled cluster with single, double and perturbative triple excitations [CCSD(T)]. We apply this protocol to study the Ov in the bulk and surface planes of rutile TiO\textsubscript{2} and rocksalt MgO, producing the first accurate and well-converged determinations of $E_\textrm{Ov}$ with this method. These reference values are used to benchmark exchange-correlation functionals in DFT and we find that all studied functionals underestimate $E_\textrm{Ov}$, with the average error decreasing along the rungs of Jacob's ladder. This protocol is automatable for high-throughput calculations and can be generalized to study other point defects or adsorbates.
\end{abstract}

\maketitle

\section{\label{sec:level1} Introduction}

Metal-oxides are a class of material with wide applications in fuel~\cite{adlerFactorsGoverningOxygen2004} and solar cells~\cite{gratzelPhotoelectrochemicalCells2001}, high-$k$ dielectrics~\cite{wangHighkGateDielectrics2018}, and the catalysis industry~\cite{ruizpuigdollersIncreasingOxideReducibility2017,asahiVisibleLightPhotocatalysisNitrogenDoped2001}. As the most prevalent defect in metal-oxides, controlling the concentration of O vacancies (Ovs) in these systems underpins much of the major advances to their applications. The dominant quantity determining the Ov concentration is the Ov formation energy, $E_\textrm{Ov}$. 

It is pivotal that $E_\textrm{Ov}$ can be determined accurately. The Ov concentration can change by several orders of magnitude with small ($\sim 0.1$ eV) changes in the $E_\textrm{Ov}$ at a given temperature~\cite{kimQuickstartGuideFirstprinciples2020}, resulting in drastic changes in thermodynamic, electronic and optical properties of the metal-oxide. For example, such changes in the Ov concentration can change an insulating oxide into a photocatalyst~\cite{panDefectiveTiO2Oxygen2013} or metal~\cite{leeIsostructuralMetalinsulatorTransition2018}. $E_\textrm{Ov}$ is also a measure of the reducibility of a metal-oxide system, with (thermal) reduction being a vital step in the thermochemical cycles for H\textsubscript{2}O and CO\textsubscript{2} splitting~\cite{meredigFirstprinciplesThermodynamicFramework2009,demlOxideEnthalpyFormation2014}. Similarly, $E_\textrm{Ov}$ has also been shown to correlate well with key catalytic properties such as the adsorption and bond activation energies of various molecules~\cite{hinumaDensityFunctionalTheory2018,kumarCorrelationMethaneActivation2016,suTrendsBondScission2020} on metal-oxide surfaces.

The reliable experimental determination of $E_\textrm{Ov}$ is very challenging as it depends sensitively on many factors~\cite{gunkelOxygenVacanciesVisible2020} (e.g.\ Ov concentration, presence of dopants, and crystallite size). As such, $E_\textrm{Ov}$ has been mostly computed through electronic structure modeling, particularly with density functional theory (DFT). However, $E_\textrm{Ov}$ is highly sensitive to the approximation of the exchange-correlation (XC) functional in DFT, oftentimes leading to large disagreements in their predictions. For example, in the $(110)$ rutile TiO\textsubscript{2} surface~\cite{ganduglia-pirovanoOxygenVacanciesTransition2007}, $E_\textrm{Ov}$ can vary by more than 1.5 eV -- around 50\% of the absolute $E_\textrm{Ov}$ -- with similar discrepancies observed in MgO~\cite{richterConcentrationVacanciesMetalOxide2013} and other metal-oxide systems~\cite{ganduglia-pirovanoOxygenVacanciesTransition2007,wellingtonOxygenVacancyFormation2018}. Additionally, DFT with the generalized gradient approximation (GGA) XC functional has been shown to be inadequate for modeling Ovs in transition metal oxides such as rutile TiO\textsubscript{2}, predicting that the unpaired electrons produced during Ov formation are delocalized~\cite{divalentinElectronicStructureDefect2006}, whilst hybrid functionals which incorporate exact exchange predict localized electrons on adjacent Ti sites~\cite{shibuyaSystematicStudyPolarons2012}. There is a clear need for high-accuracy and well-converged reference values of the $E_\textrm{Ov}$ for these systems. 

The (electrostatic) embedded cluster approach~\cite{bernsteinHybridAtomisticSimulation2009} offers the potential to efficiently apply accurate methods to study the Ov. It limits explicit quantum-mechanical calculations to only a finite-sized quantum cluster, with the electrostatic interactions from the rest of the system approximated by point charges. Over the past few decades, there have been numerous applications of this approach to the Ov in metal-oxides~\cite{pacchioniModelingDopedDefective2008}. These studies have ranged from applying the basic electrostatic embedding that has been described~\cite{bredowElectronicStructureIsolated2002,carrascoFirstprinciplesStudyOptical2005,ammalModelingNobleMetal2010}, to more involved setups which utilize polarizable environments~\cite{sushkoRelativeEnergiesSurface2000,bergerFirstprinciplesEmbeddedclusterCalculations2015,nasluzovElasticPolarizableEnvironment2003,neymanSingleDMetalAtoms2005} (of varying degrees of sophistication) or couples the quantum cluster to the environment self-consistently via the ``perturbed cluster'' approach~\cite{scorzaOxygenVacancySurface1997}. Regardless of the approach taken, the outstanding challenge is finding a quantum cluster that is sufficiently small, or even the smallest, which allows for inexpensive modeling at the reference level of theory whilst minimizing finite-size errors to converge results to the bulk (i.e.\ infinite size) limit. This process is difficult because the quantum cluster can take any chemical formula (i.e.\ size) and for each chemical formula, there can be many possible shapes, with widely differing convergence behaviors~\cite{xuClusterModelingMetal1999,luConvergenceClustersBulk1999}.

To circumvent the complexities with searching the entire size and shape space, converged clusters are normally selected from a set of clusters generated from chosen design rubrics~\cite{ammalModelingNobleMetal2010,kickTransferableDesignSolidstate2019}. The most common rubrics are to keep quantum clusters stoichiometric and spherical~\cite{scanlonBandAlignmentRutile2013,richterConcentrationVacanciesMetalOxide2013}. Identifying suitable quantum clusters which follow these design principles require significant time investment and manual input, paired with lots of trial-and-error. As such, only a handful of clusters are normally created, making it difficult to affirm the quality of a cluster as well as study multiple crystal systems simultaneously. Additionally, recent work~\cite{kickTransferableDesignSolidstate2019,dittmerComputationNMRShielding2020} has shown that these rubrics lead to poor convergence with cluster size.

There is growing interest in approaches which can lessen the manual labor involved. These approaches range from building clusters using layers (based on unit cell multipoles~\cite{neeseORCAQuantumChemistry2020} or coordination spheres~\cite{krasikovInitioEmbeddedCluster2009}) to using building blocks of either the unit cell~\cite{dittmerAccurateBandGap2019,dittmerComputationNMRShielding2020,kadossovEffectSurroundingPoint2007,luConvergenceClustersBulk1999} or fully-coordinated ions~\cite{dittmerAccurateBandGap2019,bagusClusterEmbeddingIonic2019}. Whilst the quantum cluster series generated from these approaches have been shown to converge properties of an ideal crystal (e.g.\ NMR constants~\cite{dittmerComputationNMRShielding2020}, bandgaps~\cite{dittmerAccurateBandGap2019} and optical spectra~\cite{liaoOpticalExcitationsHematite2011,bagusClusterEmbeddingIonic2019}), their extension to the study of surfaces or point defects such as the Ov is not clear and they still suffer from important deficiencies. For example, the positioning of the building blocks in the building block approach still requires manual definition. With the layered approach, the number of atoms within each layer increases significantly compared to the previous, providing little granularity in the sizes sampled. This leads to large converged quantum clusters that are not conducive for high accuracy calculations, which typically exhibit steep system size scaling.

To summarize, a systematic and general approach for designing a quantum cluster set which provides good granularity in  the sizes and shapes being sampled whilst ensuring rapid convergence is currently lacking. In this work, we propose a quantum cluster design protocol (named SKZCAM after the authors' initials) which achieves these qualities for computing accurate Ov formation energies in metal-oxides. The core enabling development is to put the control of the shape and size of a cluster into a robust and flexible framework using the radial distribution function (RDF) to divide metal cations into ``shells''. The O anion positions for each cluster arise naturally from the metal cation positions based on the criteria that all dangling bonds are removed from the metal cations. The result is a process that requires no manual intervention to generate clusters of high granularity for virtually any metal-oxide crystal system and surface termination, whilst converging rapidly with size. 

This protocol is used to obtain small converged clusters for studying the $E_\textrm{Ov}$ in the bulk and common surface planes of rocksalt MgO and rutile TiO\textsubscript{2}, two technologically relevant metal-oxide systems with properties that depend sensitively on the Ov. These clusters (involving fewer than 600 correlated electrons for all systems) enable  the accurate
local natural orbital (LNO-)CCSD(T)~\cite{rolikGeneralorderLocalCoupledcluster2011,rolikEfficientLinearscalingCCSD2013,nagyOptimizationLinearscalingLocal2017,nagyOptimizationLinearScalingLocal2018,nagyApproachingBasisSet2019} method to be applied to compute $E_\textrm{Ov}$. 
We use our computed values to benchmark common DFT XC functionals, ranging from GGAs up to double-hybrids, and find that all studied XC functionals underestimate $E_\textrm{Ov}$, with this error decreasing, on average, along the rungs of Jacob's ladder.

\section{\label{sec2} Computational Details}
We describe the embedded cluster and supercell approaches used to generate the necessary structures to evaluate $E_\textrm{Ov}$ for bulk rocksalt MgO and its $(100)$ surface as well as bulk rutile TiO\textsubscript{2} and its $(110)$ surface in this section. This quantity is calculated as:
\begin{equation} \label{neweov}
E_\textrm{Ov} = E[\textrm{D-MO}] - E[\textrm{P-MO}] + E[\textrm{O}] - \frac{1}{2}E_\textrm{bind},
\end{equation}
where $E[\textrm{P-MO}]$ and $E[\textrm{D-MO}]$ are the total energies of the pristine and defected metal-oxide systems respectively. The defected system is created from the pristine system by removing an O atom without further geometrical optimization. Both systems are treated in the closed-shell singlet state. $E[\textrm{O}]$ is the total energy of an O atom in the (unrestricted) triplet state and $E_\textrm{bind}$ is the molecular binding energy of an O\textsubscript{2} molecule, which has been computed for various levels of theory in Sec.\ S1 of the supplementary material. The final two terms sum up to (half) the energy of an O\textsubscript{2} molecule, thus defining $E_\textrm{Ov}$ under O-rich conditions. 

The definition given in Eq.~\ref{neweov} computes the \textit{unrelaxed} O vacancy formation energy $E_\textrm{Ov}$. Whilst the use of unrelaxed Ov structures precludes direct comparison with experiment when relaxation effects are significant, it provides an upper bound to the relaxed $E_\textrm{Ov}$ and a valid reference when comparing DFT values to high accuracy methods at the same geometry. In any case, relaxation effects have been found to be negligible for rocksalt MgO~\cite{orlandoCatalyticPropertiesFcentres1997,ertekinPointdefectOpticalTransitions2013}. Additionally, whilst this effect is significant for rutile TiO\textsubscript{2}~\cite{helaliScalingReducibilityMetal2017}, it is hard to quantify accurately due to a strong dependence on the spin-state (see Sec.~\ref{sec:relax_effects}) and chosen DFT XC functional (see Sec.\ S2 of the supplementary material), making it most appropriate to evaluate the \textit{unrelaxed} $E_\textrm{Ov}$ for the purposes of this work.

\subsection{Electrostatic embedded cluster calculations}
The electrostatic embedded cluster approach, illustrated in Fig.~\ref{fig_01} (a), used in this study features a quantum(-mechanical) cluster centred around the O vacancy. To model the long-range electrostatic potential from the rest of the material, this cluster is surrounded by a sphere and hemisphere of point charges of radius 30 and 40 \AA{} for the bulk and surface respectively; formal point charges have been placed at the metal and O ion crystallographic positions.
In the vicinity of the quantum cluster (< 7 \AA{}), the positive point charges are ``capped'' with the effective core potential (ECP) of the corresponding metal ion, taken from the Stuttgart/Cologne group~\cite{fuentealbaPseudopotentialCalculationsAlkalineearth1985,dolgEnergyAdjustedInitio1987}, to avoid electron leakage from the bonded O ions at the boundary of the quantum cluster. The chosen radii of the different regions (see Sec.\ S3 of the supplementary material) can converge $E_\textrm{Ov}$ to 0.01 eV. The placement of the point charges and ECPs alongside the atoms of the quantum cluster were constructed using py-ChemShell~\cite{luOpenSourcePythonBasedRedevelopment2019}.

DFT calculations were performed in ORCA~\cite{neeseORCAQuantumChemistry2020} version 5.0 and MRCC~\cite{kallayMRCCProgramSystem2020} 2020, with the latter interfaced to libXC~\cite{lehtolaRecentDevelopmentsLibxc2018}. The def2-SVP, def2-TZVPP and def2-QZVPP Weigend-Ahlrichs~\cite{weigendBalancedBasisSets2005} basis sets were used throughout this paper, with the standard def2-JK~\cite{weigendHartreeFockExchange2008,weigendAccurateCoulombfittingBasis2006} fitting basis set used for Coulomb and exchange integrals. Convergence tests indicate that the def2-SVP and def2-TZVPP basis sets exhibit errors of 0.4 and 0.02 eV w.r.t.\ the def2-QZVPP basis set (see Sec.\ S4 of the supplementary material).

Localized orbital correlated wave-function theory (cWFT) calculations were performed with the LNO-CCSD(T) and local M\o{}ller Plesset perturbation theory (LMP2) implementations of Nagy \etal{}~\cite{nagyIntegralDirectLinearScalingSecondOrder2016,nagyOptimizationLinearScalingLocal2018} in MRCC, using the ``Normal'' LNO thresholds. The MP2 contribution to the B2PLYP~\cite{grimmeDoublehybridDensityFunctional2007} XC functional was also evaluated using the LMP2 implementation of MRCC.
Complete basis set (CBS) extrapolation parameters for the def2-TZVPP and def2-QZVPP pair, CBS(TZVPP/QZVPP), taken from Neese and Valeev~\cite{neeseRevisitingAtomicNatural2011}, were used for the Hartree-Fock (HF) and correlation energy components of the cWFT total energy. 
Oxygen basis functions were placed at the Ov site and only valence electrons were correlated. Convergence tests indicate that these settings can give accuracy to within 0.1 eV (see Sec.\ S4 B of the supplementary material). Deficiencies due to frozen-core or basis set size were accounted for through a correction computed on small tractable clusters featuring a "reduced frozen-core"~\cite{bistoniTreatingSubvalenceCorrelation2017} (e.g.\ Ne for Ti and He for Mg) and basis set involving cc-pwCV$n$Z~\cite{petersonAccurateCorrelationConsistent2002,balabanovSystematicallyConvergentBasis2005} and aug-cc-pV$n$Z~\cite{kendallElectronAffinitiesFirst1992} on the metal and O ions respectively (see Sec.\ S5 of the supplementary material), which has been extrapolated~\cite{neeseRevisitingAtomicNatural2011} for $n=$ T and Q. The def2-$n$ZVPP-RI auxiliary basis sets~\cite{weigendRIMP2OptimizedAuxiliary1998,hellwegOptimizedAccurateAuxiliary2007} were used for the local cWFT calculations with the def2 basis sets, whilst the automatic auxiliary basis functions of Stoychev \etal{}~\cite{stoychevAutomaticGenerationAuxiliary2017,lehtolaStraightforwardAccurateAutomatic2021} were generated for the correlation-consistent (cc) basis sets.

\subsection{Periodic supercell calculations} \label{supercell_calcs}
Periodic supercell calculations with DFT were performed in the Vienna \textit{Ab initio} Simulation Package (VASP)~\cite{kresseEfficiencyAbinitioTotal1996,kresseEfficientIterativeSchemes1996}. These calculations serve to define the positions for constructing embedded cluster calculations as well as to provide reference $E_\textrm{Ov}$ values. All systems used structures optimized at the R2SCAN~\cite{furnessAccurateNumericallyEfficient2020} DFT level as these agree well with experimental lattice parameter values (see Sec.\ S6 of the supplementary material). The bulk rutile TiO\textsubscript{2} calculations employed a 192 atom $(2\sqrt{2} \times 2 \sqrt{2} \times 4)$ supercell and the bulk rocksalt MgO calculations used a 64-atom $(2 \times 2 \times 2)$ supercell, both with a $(2 \times 2 \times 2)$ $\Gamma$-centred Monkhorst-Pack $k$-point sampling. The $(001)$ MgO surface and $(110)$ TiO\textsubscript{2} surface calculations both employed an asymmetric five-layer slab with the top two layers allowed to relax to form the pristine surface. A $p(2\times 4)$ and $(2 \times 2)$ supercell was used for the TiO\textsubscript{2} and MgO surfaces respectively, each computed with a $(2\times2\times1)$ Monkhorst-Pack mesh and $12$ \AA{} of vacuum. For the TiO\textsubscript{2} surface, a correction to the $p(2\times 6)$ supercell size, at the PBE~\cite{perdewGeneralizedGradientApproximation1996} level, was applied for the PBE0~\cite{adamoReliableDensityFunctional1999} hybrid DFT calculation (see Sec.\ S7 of the supplementary material). An energy cutoff of 500 eV was used for all four systems, with small core projector augmented wave (PAW) potentials on the metal cations, leaving 12 and 10 valence electrons for Ti and Mg respectively. The standard PAW potential was used on the O anion.

When constructing embedded cluster systems, the optimized bulk unit cells were repeated into supercells with dimensions larger than the embedded clusters to allow for them to be cleaved out. For the surface systems, the five-layer supercell slab was concatenated (along the surface normal direction) with an additional 25 layers (taken from the bulk) before being repeated into supercells for embedded cluster calculations.

\subsection{Spin-state of the Ov in Rutile TiO\textsubscript{2}} \label{sec:relax_effects}

The \textit{unrelaxed} Ov systems of rocksalt MgO and rutile TiO\textsubscript{2}, in both their bulk and common surface planes, all feature a well-controlled closed-shell singlet state~\cite{orlandoCatalyticPropertiesFcentres1997,ertekinPointdefectOpticalTransitions2013,janottiHybridFunctionalStudies2010,bergerFirstprinciplesEmbeddedclusterCalculations2015}. Whilst this spin state is appropriate for the MgO systems (since relaxation effects are negligible), there are significant discrepancies regarding the spin-state of the relaxed Ov structures of rutile TiO\textsubscript{2} within the literature, both from experimental and computational studies.

Experimental results from electron paramagnetic resonance (EPR) and infrared spectroscopy have detected that the excess electrons, believed to originate from Ovs, localize on Ti ions of the system, forming a deep band gap state~\cite{yimOxygenVacancyOrigin2010,hendersonInsightsPhotoexcitedElectron2003,setvinDirectViewExcess2014} and pointing towards an open-shell spin state~\cite{stonehamTrappingSelftrappingPolaron2007,yangPhotoinducedElectronParamagnetic2009}. However these experiments also conflict with the high mobilities predicted by electrical measurements~\cite{yagiElectronicConductionSlightly1996}, suggesting a shallow n-type donor~\cite{nowotnyElectricalPropertiesDefect2006} which favors delocalized electrons.

Computational studies on the Ov of rutile TiO\textsubscript{2} have indicated various possible spin-states, ranging from the closed-shell singlet~\cite{chenColorCenterSinglet2020,janottiHybridFunctionalStudies2010}, to the open-shell triplet~\cite{mattioliInitioStudyElectronic2008,bergerFirstprinciplesEmbeddedclusterCalculations2015} and singlet~\cite{deakQuantitativeTheoryOxygen2012,deskinsDistributionTi3Surface2011,helaliScalingReducibilityMetal2017} states. For the open-shell states, the formation of polarons~\cite{franchiniPolaronsMaterials2021} (i.e.\ localized electrons on Ti ions) brings additional complications. Depending on the localization site of the polarons and degree of polaronic distortion, there can be a wide range of formation energies~\cite{deskinsDistributionTi3Surface2011,chretienElectronicStructurePartially2011,shibuyaSystematicStudyPolarons2012}.

A recent benchmark study using CCSD(T) by Chen \etal{}~\cite{chenColorCenterSinglet2020} has found that the closed-shell singlet state is still the most stable state with the inclusion of relaxation effects. The validity of CCSD(T) for studying the Ov in TiO\textsubscript{2} was confirmed from preliminary full configuration interaction quantum Monte Carlo (FCIQMC) calculations, which reveal the single-reference character of the Ov state. However it should be noted that Chen \etal{} did not directly model polarons due to the large cluster sizes required~\cite{kickTransferableDesignSolidstate2019}, which could potentially change the conclusions. We expect that the advances detailed in this work may provide the possibility to resolve these open questions/discrepancies for the TiO\textsubscript{2} system in future studies.

\section{Results}

\subsection{Quantum Cluster Design Protocol} \label{sec:design_protocol}
\begin{figure*}[h]
	\includegraphics{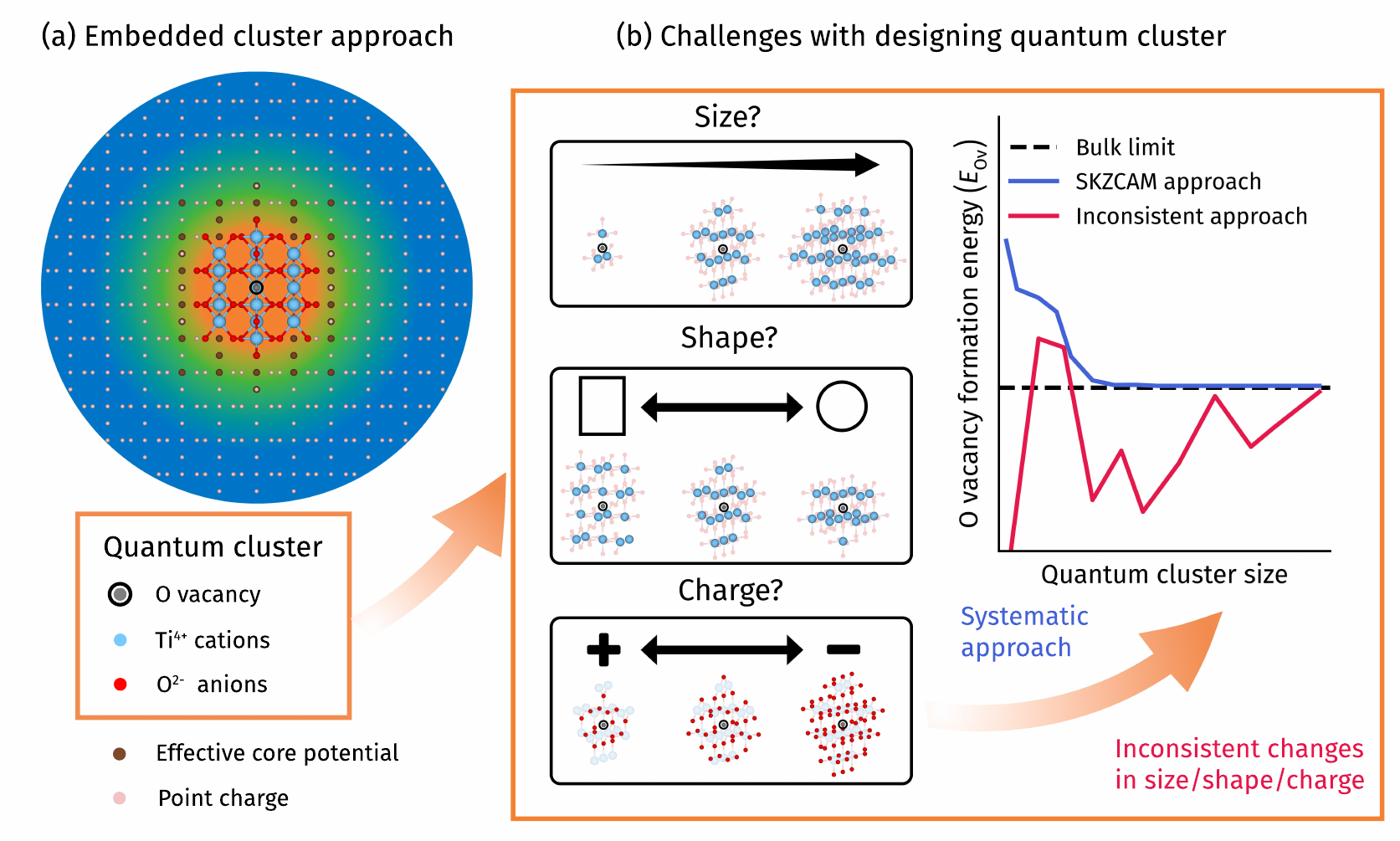}%
	\caption{\label{fig_01} (a) Schematic of the electrostatic embedding approach for bulk rutile TiO\textsubscript{2}. The quantum cluster (in the orange region) is treated with the electronic structure theory of choice. It is surrounded by a spherical (hemispherical) field of point charges in the green and blue regions for a bulk (surface) system. In the vicinity (green region) of the quantum cluster, the point charges are replaced with effective core potentials to prevent spurious charge leakage out of the quantum cluster. Normally, converged quantum clusters are selected from a series of clusters constructed through chosen design rubrics. Panel (b) highlights the key challenges with designing such a series of quantum clusters for metal-oxides: (i) deciding what sizes (in terms of number of metal cations, visualized as light blue spheres) to sample; (ii) deciding how to arrange these metal cations (i.e.\ shape); (iii) and choosing the charge of each quantum cluster, which is largely controlled by the number of O anions (visualized as red spheres). Depending on the chosen rubrics for these three factors, convergence of the O vacancy formation energy $E_\textrm{Ov}$ can be either systematic or inconsistent, as indicated by the blue and red lines in the schematic graph. The points representing the `Inconsistent approach' were taken from calculations of stoichiometric, spherical-shaped quantum clusters with inconsistent changes in the number of Ti ions (i.e.\ size) and spatial arrangement of O anions (i.e.\ charge).}
\end{figure*}

Designing a systematic and general set of quantum clusters for metal-oxides is a complex process; we have identified and summarized the key challenges involved in Fig.~\ref{fig_01} (b). Two highly important and interdependent factors are the size and shape of the quantum cluster. For metal-oxides, these two factors are predominantly controlled by the metal cations in the quantum cluster since these systems can be considered as formed from polyhedrons (e.g.\ TiO\textsubscript{6} octahedron for TiO\textsubscript{2}) centered around the metal cations -- highlighted in light blue for bulk rutile TiO\textsubscript{2} in Fig.~\ref{fig_01} (b). The choice of studied quantum cluster sizes is normally a manual process, often involving large arbitrary changes in sizes along the series. It is important that this process can be made systematic, whilst sampling clusters in small increments, so that small -- even the smallest -- converged clusters can be found for the property being studied, particularly when applying expensive cWFT methods. For a given size, the metal cations can take up virtually any spatial arrangement (i.e.\ shape) and $E_\textrm{Ov}$ can vary significantly depending on the chosen shape, as shown in Sec.\ S8 A of the supplementary material. Under the electrostatic embedding scheme, the quantum cluster can be chosen to take up any charge. 
Thus, for a given metal cation configuration (i.e.\ size and shape), there can be virtually any number of O anions; if the ratio of O to Ti ions is more (less) than the stoichiometric ratio, then a negatively (positively) charged quantum cluster is formed. 
Here, the same challenges with deciding the number and spatial arrangement of O anions exist as the metal cations. If any of the factors discussed above are performed in an inconsistent manner, the convergence of the quantum cluster series becomes non-monotonic as shown by the red line of the schematic graph in Fig.~\ref{fig_01} (b). 
In this work, we propose the SKZCAM approach for constructing a set of quantum clusters which shows fast and systematic convergence towards the bulk limit (blue line in the schematic graph of Fig.~\ref{fig_01} (b)). 
We define a rigorous framework for deciding the metal cation configurations in the quantum cluster series, with the O anion configuration naturally arising based on a separate robust rubric.

\begin{figure*}
	\includegraphics{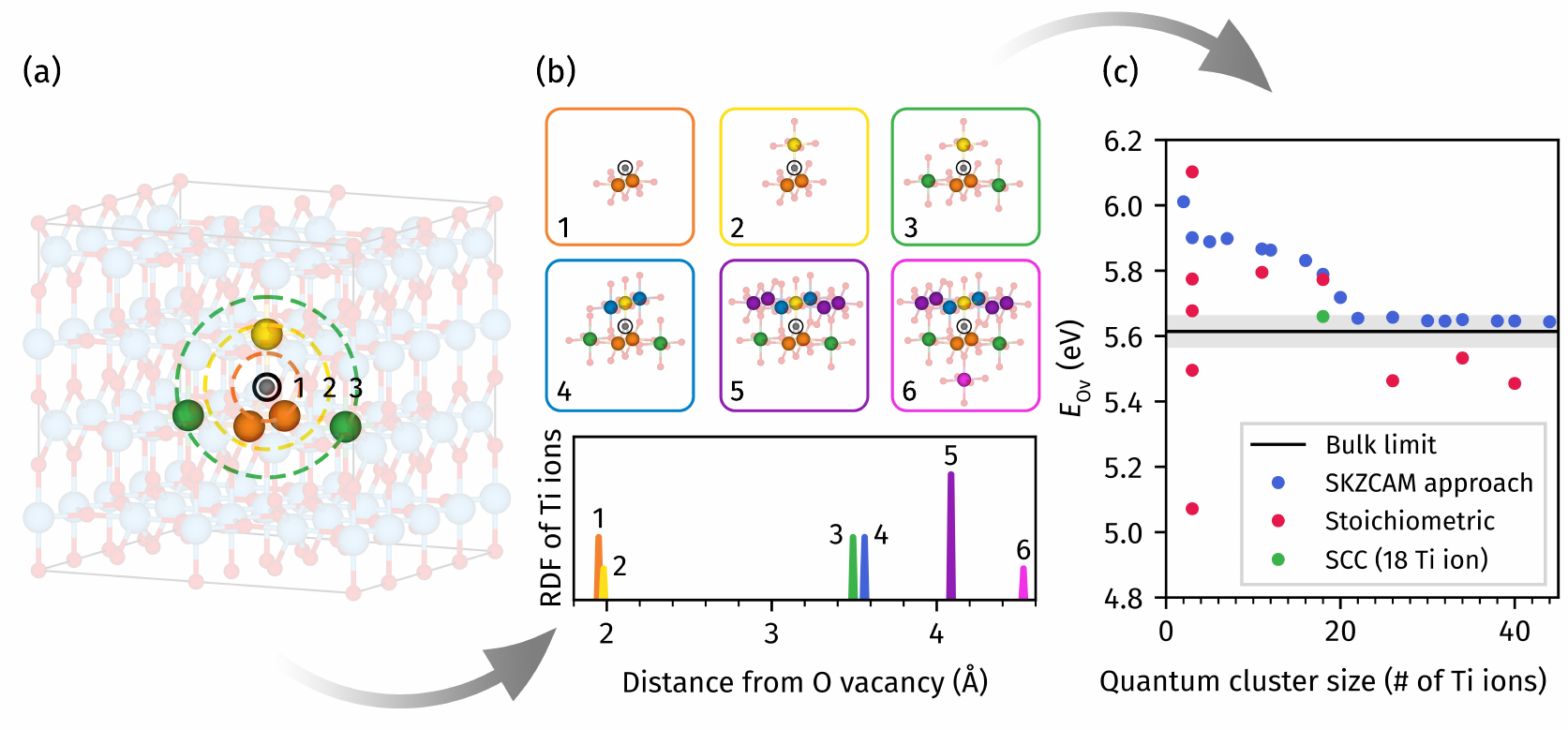}
	\caption{\label{fig_02} Illustration of the SKZCAM approach for designing the quantum clusters described in this work, using bulk rutile TiO\textsubscript{2} as an example. Around the gray O vacancy (Ov) outlined by a black circle, the Ti cations arrange as ``shells'' consisting of equidistant, symmetry related cations; we highlight the first three shells in (a) with unique colors. These shells appear as peaks in the radial distribution function (RDF) $\left [ 4 \pi r^2g(r) \right ]$ plot of Ti cations around the Ov in (b). Starting from the first shell, clusters of systematically increasing size can be generated by incorporating subsequent shells. We give the example of the first six quantum clusters generated in (b). In our approach, we choose O anions, illustrated by translucent red spheres, such that all dangling bonds around the Ti cations are removed. This combination leads to a systematic convergence of $E_\textrm{Ov}$ at the PBE-DFT level towards the bulk limit with quantum cluster size, as indicated by the blue markers in (c). The red markers correspond to stoichiometric clusters created to have the same Ti cations as a subset of the SKZCAM approach clusters, but involving fewer O anions to meet the stoichiometric ratio. We observe poor non-monotonic convergence towards the bulk limit with this choice. 
	The bulk limit in (c) was calculated from a supercell calculation described in Sec.~\ref{supercell_calcs}, whilst embedded cluster calculations were performed with the def2-TZVPP basis set.}
\end{figure*}

The RDF of the number of metal cations as a function of distance from the Ov is used to generate the metal cation configurations in the quantum cluster series. Using bulk rutile TiO\textsubscript{2} as an example in Fig.~\ref{fig_02} (a), we see that the metal Ti cations arrange as shells (with the first three given distinct colors) of symmetry-related equidistant cations around the Ov, denoted by the gray sphere within a black circle. In an RDF plot, given at the bottom panel of Fig.~\ref{fig_02} (b), these Ti cation shells will appear as distinct peaks. Starting from the first metal cation shell/peak found in the RDF plot, metal cation configurations of systematically increasing size can be created by adding subsequent metal cation shells, with the first six metal cation configurations for this series visualized in Fig.~\ref{fig_02} (b). The RDF shells/peaks are completely controlled by the crystal structure, point defect site and surface termination of the system, requiring no manual input. Furthermore, it provides good granularity in the size (i.e.\ number) of metal cations sampled in the quantum cluster series. At the same time, a variety of shapes are studied, ranging from spherical to cuboidal, all whilst ensuring symmetry about the point defect is maintained.

The metal cations serve as the base for the subsequent O anion configuration in each cluster. It is most physical to consider only O sites which ensure at least one bond to a nearby metal cation, as shown as translucent red spheres for the quantum clusters in Fig.~\ref{fig_02} (b). In principle, any number of these O anion sites can be used; we take the unambiguous approach of placing O anions in all shown positions, ensuring no dangling bonds on the metal cations. This is the most chemically intuitive approach because it is equivalent to fully-coordinating all of the metal cations for bulk systems. For surface systems, some metal cations at the surface are not fully-coordinated due to the nature of the surface termination. As a result, the O anion configuration is completely determined by the metal cation configuration, which is in turn controlled by the RDF of metal cations around the point defect. We note that because the ratio of O anions to metal cations exceeds the stoichiometric ratio with this choice, the resulting quantum clusters are all negatively charged. 

Beyond being fully systematic and general whilst providing good granularity, this SKZCAM approach also converges rapidly with cluster size. As shown by the blue markers in Fig.~\ref{fig_02} (c), $E_\textrm{Ov}$ becomes converged to within 0.05 eV of the bulk limit (illustrated by the gray error bars) for the 22 Ti ion cluster (consisting of the first ten RDF shells/peaks), with all subsequent clusters staying converged. As comparison, we have also constructed stoichiometric neutral clusters for a subset of the Ti cation configurations as shown by the red markers. These clusters are the most common type within the literature and we see a poor non-monotonic convergence towards the bulk limit, requiring clusters larger than the studied range of sizes. Notwithstanding the slow convergence, there is also significant ambiguity in constructing stoichiometric clusters for a given metal cation configuration since there are more possible O anion sites than allowed by stoichiometry. We give the $E_\textrm{Ov}$ for 6 possible O anion configurations in the 3 Ti ion quantum cluster (visualized in Sec.\ S8 B) in Fig.~\ref{fig_02} (c) - there is a wide range in $E_\textrm{Ov}$ of 1.0 eV. For larger (stoichiometric) cluster sizes, there will be more possible O anion configurations; we have calculated $E_\textrm{Ov}$ for only one such O anion configuration at larger cluster sizes based on including O anions closest to the Ov.

Whilst the converged 22 Ti ion cluster found is already quite small, it may be beneficial to search for smaller converged clusters, particularly for performing expensive cWFT calculations. The SKZCAM approach provides a robust framework for defining the shape of a quantum cluster based on metal cation shells, which serve as the building blocks of the cluster. The 22 Ti ion cluster consists of the first ten metal cation RDF shells/peak, with the removal of the furthest (tenth) shell leading to an unconverged 20 Ti ion cluster. By systematically removing closer RDF shells from the 22 Ti ion cluster (see Sec.\ S9 of the supplementary material), we identify a smaller 18 Ti ion cluster formed from removing the seventh metal cation shell (see green marker in Fig.~\ref{fig_02} (c)). Removal of any subsequent shells from this 18 Ti ion cluster leads to large changes in $E_\textrm{Ov}$.

\subsection{Converged Clusters for the Ov}

\begin{figure*}
	\includegraphics{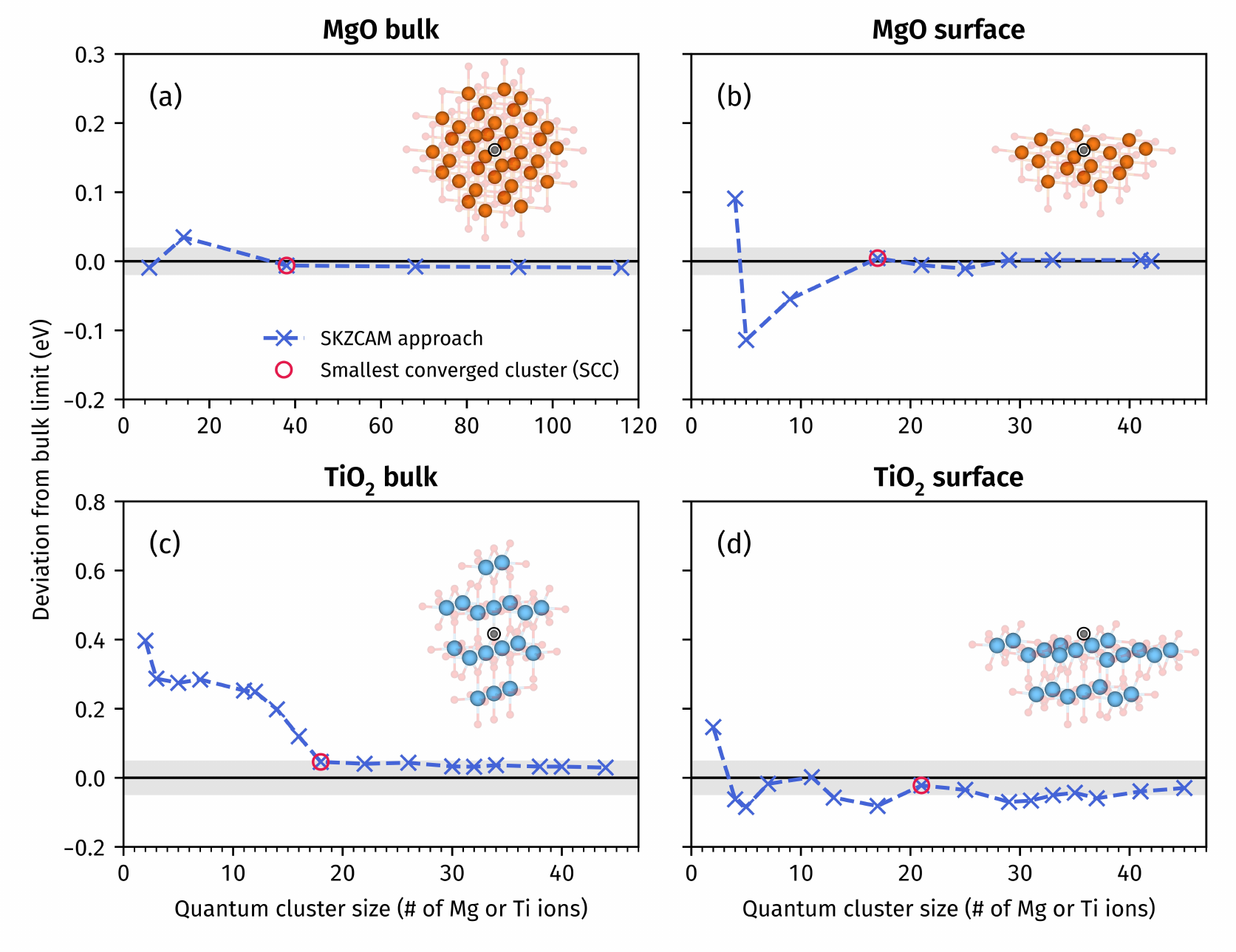}%
	\caption{\label{fig_03} The deviation from the bulk limit of the O vacancy formation energy ($E_\textrm{Ov}$) as a function of quantum cluster size -- generated via the SKZCAM approach -- for (a) MgO bulk, (b) MgO surface, (c) TiO\textsubscript{2} bulk and (d) TiO\textsubscript{2} surface. The smallest converged cluster is marked with a red circle for each system and illustrated in each panel, where the light blue, orange and translucent red spheres correspond to Ti, Mg and O ions respectively. For the MgO systems, embedded cluster calculations were performed at the def2-SVP PBE-DFT level, with a correction to the def2-TZVPP basis to enable comparison to the bulk limit (see Sec.\ S10 of the supplementary material). The TiO\textsubscript{2} systems feature explicit def2-TZVPP calculations (to ensure smoother convergence with cluster size), with PBE and PBE0 used for the bulk and surface respectively. The bulk limit results were calculated using the supercell approach with the corresponding DFT level in each panel.}
\end{figure*}

The SKZCAM approach outlined in the previous section is applied to create systematic sets of quantum clusters to study $E_\textrm{Ov}$ at the DFT level and beyond for bulk rocksalt MgO and its $(001)$ surface, as well as bulk rutile TiO\textsubscript{2} and its $(110)$ surface in Fig.~\ref{fig_03}. 

For the MgO systems and TiO\textsubscript{2} bulk, there is systematic convergence towards the bulk limit, approximated from a supercell calculation. We consider convergence reached at the point when $E_\textrm{Ov}$ starts to plateau, with the cluster located at this point being the smallest converged cluster (SCC). We find an SCC with 38 Mg ions and 17 Mg ions for MgO bulk and surface respectively. Beyond these SCC sizes, the $E_\textrm{Ov}$ of larger clusters are less than 0.02 eV from the bulk limit, as illustrated by the gray error bars. Although the 6 Mg ion quantum cluster for bulk MgO is within 0.02 eV of the bulk limit, this agreement is fortuitous at the DFT level because at higher levels of theory (LMP2), it differs by > 0.3 eV from larger converged cluster sizes (see Sec.\ S11 A of the supplementary material). For TiO\textsubscript{2} bulk, the SCC with 18 Ti ions shows an error of 0.05 eV w.r.t.\ the bulk limit, with this error decreasing slowly with cluster size subsequently. This slow convergence could arise because the electrostatic embedding setup used here does not explicitly include (long-range) polarization effects; more sophisticated setups~\cite{bergerEmbeddedclusterCalculationsNumeric2014,kubasSurfaceAdsorptionEnergetics2016,buckeridgePolymorphEngineeringTiO22015} including polarizable force fields beyond the quantum cluster could potentially improve this convergence.

The series of quantum clusters for the MgO systems were generated using the approach described in Section III of adding RDF shells of increasing distance from the Ov. We do not expect for there to be smaller converged clusters (in terms of number of Mg cations) beyond the sampled series for these systems due to their high degree of ionic bonding and cubic symmetry. For rutile TiO\textsubscript{2}, where there is a degree of directional covalent bonding~\cite{kochChargeStateTitanium2017} and anisotropy, there may be even smaller clusters. As discussed in Sec.~\ref{sec:design_protocol}, the SKZCAM approach provides the flexibility to find these smaller converged clusters, as has been done to find the 18 Ti ion SCC in TiO\textsubscript{2} bulk.

The $(110)$ rutile TiO\textsubscript{2} surface requires a separate discussion due to its complex electronic structure. As seen in Sec.\ S12 of the supplementary material, the ``noisy'' nature of its convergence arises from well-behaved odd-even oscillations in $E_\textrm{Ov}$ when Ti ions are added along specific crystallographic directions of the surface. Such odd-even size oscillations commonly appear in rutile TiO\textsubscript{2} surface~\cite{bredowElectronicPropertiesRutile2004,hameeuwRutileTiO21102006,liuStructureDynamicsLiquid2010} calculations. We find that the amplitude of these oscillations are correlated to the degree of self-interaction error in the electronic structure method, being weaker in PBE0 compared to PBE, and completely absent in methods (e.g.\ HF and LMP2) that do not suffer from self-interaction error. The 21 Ti ion cluster was selected as the SCC on the basis of good agreement with the bulk limit at the PBE and PBE0 levels (< 0.03 eV for the latter) as well as having reached the convergence plateau for HF and MP2 (see Sec.\ S11 B of the supplementary material). The gray 0.05 eV error bar in Fig.~\ref{fig_03} (d) indicates the average error of the clusters larger than (and including) the chosen SCC.

\begin{figure}
	\includegraphics{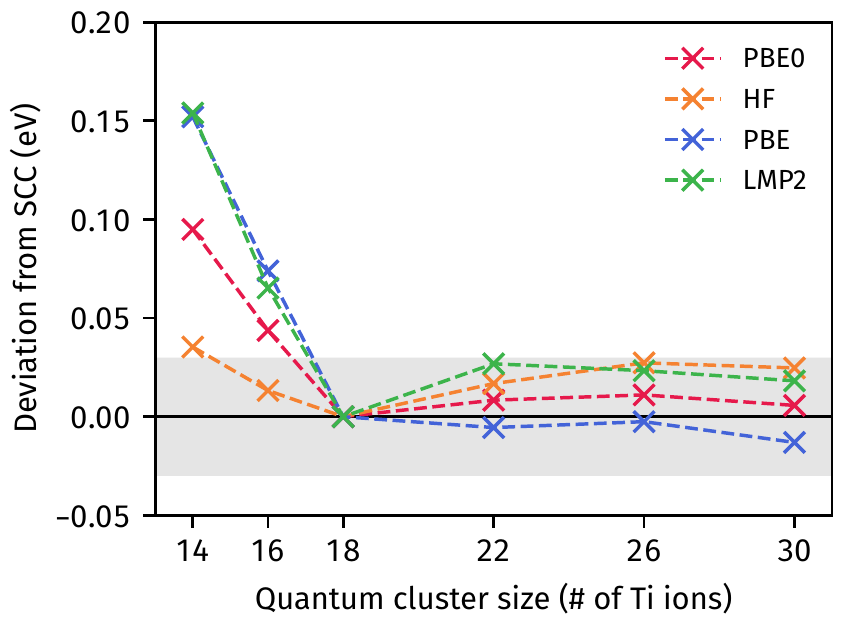}%
	\caption{\label{fig_04} The variation in the O vacancy formation energy ($E_\textrm{Ov}$) with cluster size around the smallest converged cluster (SCC), containing 18 Ti ions, for bulk rutile TiO\textsubscript{2} at the PBE, PBE0, HF and LMP2 levels of theory. The gray 0.03 eV error bar indicates the maximum level of deviation observed across the three levels of theory for sizes beyond the SCC. All calculations were performed with the def2-TZVPP basis set.}
\end{figure}

We find evidence that the SCC determined at an appropriate DFT level also leads to converged clusters -- not necessarily the smallest possible -- at other levels of theory, including cWFT methods. As seen in Fig.~\ref{fig_04} for bulk rutile TiO\textsubscript{2}, the SCC (predicted from PBE calculations) shows small changes that are less than 0.03 eV (indicated by the gray error bars) in $E_\textrm{Ov}$ compared to larger quantum clusters at all studied levels of theory, from HF to PBE0 and LMP2. We observe the same behavior in the MgO systems and TiO\textsubscript{2} surface (see Sec.\ S11 of the supplementary material).

Whilst there is good agreement between all levels of theory in the quantum clusters larger than (and including) the SCC from the DFT calculation, their behavior can vary significantly for smaller unconverged clusters. To bypass the steep scaling of cWFT methods, it is common to employ the $\Delta^\textrm{HL}_\textrm{LL}$ approach within the literature~\cite{boeseAccurateAdsorptionEnergies2013,richterConcentrationVacanciesMetalOxide2013,tsatsoulisComparisonQuantumChemistry2017} to produce reference quantities. Here, the difference between a high-level (HL) and a low-level (LL) theory (e.g.\ DFT) for a small tractable quantum cluster is added as a correction to the same low level theory computed at the bulk and basis set limits. The implicit underlying assumption is that the difference between the HL and LL methods stays the same regardless of cluster size. The differing size convergence of the various levels of theory suggests that such an assumption could exhibit uncontrolled errors if performed for arbitrarily small crystals. 

\subsection{Reference Ov Formation Energy} \label{reference_val}
Recent advances in localized orbital variants of CCSD(T) (e.g.\ LNO-CCSD(T)~\cite{nagyApproachingBasisSet2019}, DLPNO-CCSD(T)~\cite{riplingerEfficientLinearScaling2013}, PNO-LCCSD(T)~\cite{maExplicitlyCorrelatedLocal2018}, etc.) have extended the remit of coupled cluster methods. To put these advances into perspective, for the def2-QZVPP basis set on a node equipped with 72 CPU cores, a PBE and PBE0 single-point calculation on the 17 Mg ion SCC for MgO surface took 1.5 and 2.5 hours respectively, whilst this time increased to only 7 hours for LNO-CCSD(T). Such a calculation would be far outside the reach of canonical CCSD(T), with the largest studied being a cluster with 6 Mg ions and 9 O anions~\cite{richterConcentrationVacanciesMetalOxide2013}. These developments, alongside the identification of relatively small and converged clusters, enable us to compute Ov at the local CCSD(T) level with large basis sets.  The O vacancy formation energy, $E_\textrm{Ov}$, computed with LNO-CCSD(T) in the O-rich limit for the four systems are $7.68 \pm 0.15$ eV (MgO bulk), $7.18 \pm 0.15$ eV (MgO $(001)$ surface), $6.39 \pm 0.15$ eV (TiO\textsubscript{2} bulk) and $5.55 \pm 0.15$ eV (TiO\textsubscript{2} $(110)$ surface). The decrease in $E_\textrm{Ov}$ moving from bulk to surface can be expected due to the lowered coordination around the O anion sites on the surface. 

Error bars have been added to the LNO-CCSD(T) values reported above. These have been conservatively estimated at 0.15 eV. This estimate comprises of errors arising due to: (i) basis set size and frozen core treatment; (ii) local approximation thresholds; and (iii) quantum cluster finite size errors. Tests on smaller clusters (see Sec.\ S4 B of the supplementary material) shows that the (i) CBS(TZVPP/QZVPP) basis and frozen core treatment (Ar on Ti and Ne on Mg) chosen give good agreement ($\sim 0.1$ eV) w.r.t.\ a larger basis set and small frozen core (Ne on Ti and He on Mg). Based on the deviations observed in small clusters, our best estimates of $E_\textrm{Ov}$ were corrected for the bias due to the basis set and frozen core treatment (see Sec.\ S5 of the supplementary material). The (ii) LNO threshold settings chosen have been validated against canonical CCSD(T) to give small errors of < 0.04 eV (see Sec.\ S13 of supplementary). Finite size errors are expected to be less than 0.05 eV -- the typical error found for DFT w.r.t.\ the bulk limit (in Fig.~\ref{fig_03}) as well as from LMP2 calculations against larger clusters (see Sec.\ S11 of the supplementary material). Given the fast basis set convergence and all-electron nature of the embedded cluster DFT calculations, finite size errors (0.05 eV) are expected to be the dominant source of error.

\subsection{Comparison of DFT Functionals}

\begin{figure}
	\includegraphics{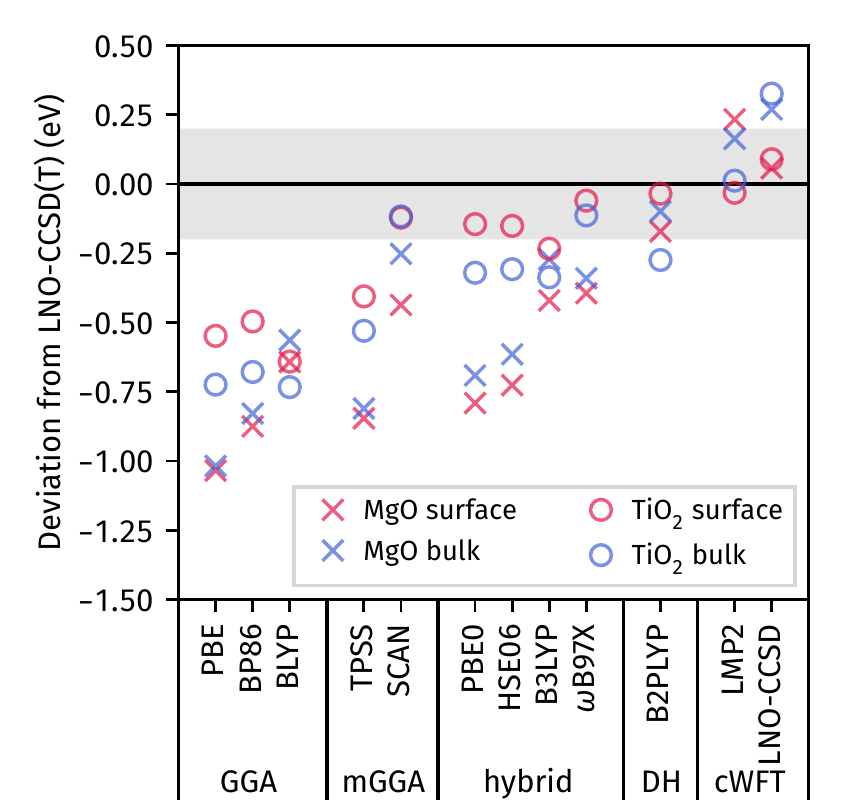}%
	\caption{\label{fig_05} Comparison of various DFT XC functionals, ranging from GGAs to double-hybrids, as well as LMP2 and LNO-CCSD to LNO-CCSD(T) reference calculations of the O vacancy formation energy ($E_\textrm{Ov}$) in MgO bulk, MgO surface, TiO\textsubscript{2} bulk and TiO\textsubscript{2} surface. DFT calculations were performed at the def2-QZVPP level with the cWFT and double-hybrid (DH) calculations following the same procedure (discussed in the text) as the LNO-CCSD(T) calculations. The gray error bars are included to indicate the combined 0.2 eV error of the DFT (0.05 eV from finite size errors) and LNO-CCSD(T) (0.15 eV as discussed in Sec.~\ref{reference_val}) values.}
\end{figure}

On top of enabling the accurate LNO-CCSD(T) method to be applied, the embedded cluster approach allows virtually all DFT XC functionals to be applied at low cost, including double-hybrids - normally computationally inaccessible for solid-state periodic DFT codes. We use our obtained LNO-CCSD(T) reference values to evaluate the performance for a range of DFT XC functionals, with their deviation from LNO-CCSD(T) $E_\textrm{Ov}$ values plotted in Fig.~\ref{fig_05}. These errors, alongside the computed $E_\textrm{Ov}$, for all of the studied methods are summarized in Table~\ref{tab:final_eov}. For all 4 systems, the studied XC functionals underestimate $E_\textrm{Ov}$ w.r.t.\ LNO-CCSD(T). This error decreases, on average, when using XC functionals on higher rungs, with meta-GGAs, hybrids and double-hybrids (DH) showing consistent improvement over the GGAs. The observed trends and variations of the XC functionals are quite consistent between the bulk and surface of the same system, but differs from one material to the next.

\begin{table*}
	\caption{\label{tab:final_eov} $E_\textrm{Ov}$ (in eV) estimates at various levels of DFT XC functional approximations for the smallest converged clusters (SCCs) of rocksalt MgO and rutile TiO\textsubscript{2}, in their bulk and common surface planes. The final estimate of the B2PLYP, LMP2, LNO-CCSD and LNO-CCSD(T) methods are also given. These values are calculated using the O\textsubscript{2} binding energy of the corresponding electronic structure level (see Sec.\ S1 of the supplementary material) in Eq.~\ref{neweov}. Errors are given w.r.t.\ the final LNO-CCSD(T) estimate. The computational details for these calculations are provided in the text.}
\begin{ruledtabular}
\begin{tabular}{lccccccccc}
            & MgO bulk &   Error    & MgO surface &   Error    & TiO\textsubscript{2} bulk &   Error    & TiO\textsubscript{2} surface &  Error      & MAE  \\ \hline
PBE         & 6.66     & -1.02 & 6.15        & -1.03 & 5.67      & -0.72 & 5.00         & -0.55 & 0.83 \\
BP86        & 6.85     & -0.83 & 6.31        & -0.87 & 5.71      & -0.68 & 5.06         & -0.50 & 0.72 \\
BLYP        & 7.11     & -0.56 & 6.54        & -0.64 & 5.66      & -0.73 & 4.91         & -0.64 & 0.64 \\
TPSS        & 6.87     & -0.81 & 6.34        & -0.85 & 5.86      & -0.53 & 5.15         & -0.41 & 0.65 \\
SCAN        & 7.43     & -0.25 & 6.74        & -0.44 & 6.27      & -0.12 & 5.43         & -0.12 & 0.23 \\
PBE0        & 6.99     & -0.69 & 6.39        & -0.79 & 6.07      & -0.32 & 5.41         & -0.14 & 0.49 \\
HSE06       & 7.06     & -0.61 & 6.46        & -0.73 & 6.08      & -0.31 & 5.40         & -0.15 & 0.45 \\
B3LYP       & 7.41     & -0.27 & 6.76        & -0.42 & 6.05      & -0.34 & 5.32         & -0.23 & 0.32 \\
$\omega$B97X& 7.34     & -0.34 & 6.79        & -0.39 & 6.28      & -0.11 & 5.49         & -0.06 & 0.23 \\
B2PLYP      & 7.58     & -0.10 & 7.01        & -0.17 & 6.12      & -0.27 & 5.52         & -0.03 & 0.14 \\
LMP2        & 7.84     & 0.16  & 7.41        & 0.23  & 6.40      & 0.01  & 5.52         & -0.03 & 0.11 \\
LNO-CCSD    & 7.95     & 0.27  & 7.24        & 0.06  & 6.72      & 0.33  & 5.64         & 0.09  & 0.19 \\
LNO-CCSD(T) & 7.68     & --    & 7.18        & --    & 6.39      & --    & 5.55         & --    &      \\
MAE         &          & 0.55  &             & 0.63  &           & 0.41  &              & 0.28  &    
\end{tabular}
\end{ruledtabular}
\end{table*}

Out of the studied DFT XC functionals, the double-hybrid B2PLYP functional shows the best agreement to the LNO-CCSD(T) reference (mean absolute error, MAE, of 0.14 eV), with all points lying within the combined DFT and LNO-CCSD(T) error bars in Fig.~\ref{fig_05}. The $\omega$B97X~\cite{chaiSystematicOptimizationLongrange2008} hybrid and SCAN~\cite{sunStronglyConstrainedAppropriately2015} meta-GGA XC functionals give the next best performance, both with an MAE of 0.23 eV. In the hybrid functionals, B3LYP~\cite{beckeDensityFunctionalThermochemistry1993} also shows good (MAE of 0.32 eV) and consistent performance across the 4 systems. On the other hand, PBE0 as well as HSE06~\cite{heydHybridFunctionalsBased2003} -- two common hybrid functionals for metal-oxide systems -- give variable performance, having large errors (of around 0.7 eV) for the MgO systems, which lowers (to around 0.2 eV) in the TiO\textsubscript{2} systems. The GGAs all severely underestimate $E_\textrm{Ov}$, with MAE's all exceeding 0.6 eV. The XC functionals in Fig.~\ref{fig_05} are arranged based on decreasing MAE within their respective Jacob's ladder rungs. XC functionals which incorporate the Becke-88 (B) exchange and Lee-Yang-Parr (LYP) correlation functionals, such as BLYP and B3LYP, are one of the top performers in their respective rungs, with B2PLYP being the best overall studied XC functional, as discussed previously. 

Lower-level cWFT methods such as LMP2 and LNO-CCSD are automatically generated in any LNO-CCSD(T) calculation and they are also compared in Fig.~\ref{fig_05}. Both LMP2 and LNO-CCSD show excellent agreement to LNO-CCSD(T), with MAEs of 0.11 and 0.19 eV respectively, and only the B2PLYP DFT XC functional has errors of a similar (small) size. As the errors of these three methods are all close to or within the error bars of the LNO-CCSD(T) values, it is not possible to ascertain which method performs better. We do find, however, that the good agreement of LMP2 appears to arise from fortuitous error cancellations, as elaborated in Sec.~\ref{sec:dft_under}.

\section{Discussion}

The two key developments of this work are: (i) a protocol for obtaining small converged clusters for performing reference calculations of the Ov at high levels of theory; and (ii) the assessment of the accuracy of XC functionals for studying the $E_\textrm{Ov}$. It is important that both these developments are properly contextualized, either from physical theory or comparison to the literature. For the latter development, we will try to rationalize the observed trends in performances of the XC functionals in Sec.~\ref{sec:dft_under}. For the former, we will compare our LNO-CCSD(T) reference values to those in the literature in Sec.~\ref{sec:mgo_bulk_compare}.

\subsection{Origin of DFT Underestimation} \label{sec:dft_under}

Seeking to understand the relative performance of DFT XC functionals in complex systems, such as metal-oxides, is not straightforward. In particular, $E_\textrm{Ov}$ is a quantity that depends on many factors. Nonetheless, we believe our results can reveal some useful insights on this property.

Systematic studies on a series of metal-oxides have shown that there is a correlation between the band gap and $E_\textrm{Ov}$ in these systems~\cite{demlIntrinsicMaterialProperties2015,helaliScalingReducibilityMetal2017}, as is confirmed by the larger $E_\textrm{Ov}$ in MgO (with a PBE band gap of 4.83 eV in the bulk~\cite{ertekinPointdefectOpticalTransitions2013}) over TiO\textsubscript{2} (1.88 eV PBE band gap~\cite{chiodoSelfenergyExcitonicEffects2010}) in our results. Physically, this correlation arises because the removal of an O atom leaves two electrons (originally occupying the O $2p$ band), which must redistribute by occupying an empty band from the conduction band~\cite{demlOxideEnthalpyFormation2014}. This redistribution energy, and in turn $E_\textrm{Ov}$, would then be correlated with the size of the band gap of the crystal system.

Within the same crystal system, we find that the predicted DFT band gaps provides a rational basis for understanding the underestimation of $E_\textrm{Ov}$ in most DFT XC functionals as well as the relative trends between the Jacob's ladder rungs. More specifically, the errors in $E_\textrm{Ov}$ w.r.t.\ reference methods should be related to the deviations of the XC functionals from the experimental or reference method band gap. Due to the presence of self-interaction error~\cite{baoSelfInteractionErrorDensity2018} (SIE), semilocal functionals (meta-GGAs and below) will underestimate the band gap, hence underestimating $E_\textrm{Ov}$, with meta-GGAs normally giving improved band gaps over GGAs~\cite{borlidoExchangecorrelationFunctionalsBand2020}. Hybrid functionals correct for some of the SIE via the incorporation of exact exchange, with the possibility that band gaps are overestimated~\cite{borlidoLargeScaleBenchmarkExchange2019,sunStronglyConstrainedAppropriately2015}. For both the rocksalt MgO and rutile TiO\textsubscript{2} systems, most of the studied hybrid functionals underestimate the band gap~\cite{dittmerAccurateBandGap2019,vinesSystematicStudyEffect2017}, hence underestimating $E_\textrm{Ov}$. 

The band gap can only serve as a general guide for the relative performance of DFT XC functionals and there are functionals which do not follow this trend, suggesting that there are other important factors which influence $E_\textrm{Ov}$. The key example is the PBE0 functional in rutile TiO\textsubscript{2}, which underestimates $E_\textrm{Ov}$ despite predicting a band gap~\cite{chiodoSelfenergyExcitonicEffects2010} of 4.05 eV that overestimates experimental electronic band gaps from photoemission experiments, typically in the range of 3.3--4.0 eV~\cite{breckenridgeElectricalPropertiesTitanium1953,tezukaPhotoemissionBremsstrahlungIsochromat1994,hardmanValencebandStructureMathrmTiO1994,ranganEnergyLevelAlignment2010}. Additionally, despite predicting band gaps that are 0.6 and 1.0 eV higher than HSE06 for bulk MgO~\cite{garzaPredictingBandGaps2016,fritschSelfConsistentHybridFunctional2017} and rutile TiO\textsubscript{2}~\cite{chiodoSelfenergyExcitonicEffects2010,janottiHybridFunctionalStudies2010} respectively, these two functionals have very similar performances (both with MAE close to 0.5 eV) across the range of systems. Despite predicting worse band gaps~\cite{zhangSubtletyTiO2Phase2019}, the improved overall performance of the SCAN functional (with MAE of 0.23 eV) over many hybrid functionals is another example.

In GGAs, on top of the underestimated band gap due to the SIE, a major contribution to its errors also arises from a poor description of the binding energy of the O\textsubscript{2} molecule -- they tend to predict strong overbinding (e.g.\ 0.50 eV/atom for PBE). It is common within the literature to correct for this error by replacing the GGA O\textsubscript{2} binding energy in Eq.~\ref{neweov} with the experimental binding energy of 5.22 eV~\cite{fellerReexaminationAtomizationEnergies1999}. With this change, the underestimation is decreased significantly, with MAE decreases of the range of 0.3--0.5 eV for the GGAs (see Sec.\ S14 of the supplementary material). XC functionals on higher rungs are less impacted, with improvements in the MAE of less than 0.2 eV for the meta-GGAs and hybrids. We also find that the better performance of LMP2 over (the more accurate) LNO-CCSD method is fortuitous, arising from its overbinding of 0.22 eV/atom (in Sec. S1 of the supplementary material) and it exhibits a larger MAE of 0.34 eV relative to the 0.11 eV of LNO-CCSD when using the experimental binding energy.

\subsection{Comparison to Previous Work for MgO Bulk} \label{sec:mgo_bulk_compare}

As the prototypical system for studying the Ov in metal-oxides, bulk MgO has been subject to several studies involving high level reference methods~\cite{ertekinPointdefectOpticalTransitions2013,richterConcentrationVacanciesMetalOxide2013,scorzaOxygenVacancySurface1997,galloPeriodicEquationofmotionCoupledcluster2021,rinkeFirstPrinciplesOpticalSpectra2012} and out of the 4 studied systems, it is the only system, to our knowledge, where $E_\textrm{Ov}$ has been experimentally determined~\cite{kappersEnsuremathCentersMagnesium1970}. Hence, bulk MgO makes for a good system to assess the accuracy of our obtained reference LNO-CCSD(T) values. 

\begin{table}[]
	\caption{\label{tab:lit_compare}The O vacancy formation energy ($E_\textrm{Ov}$) in bulk MgO obtained through high-level theory methods and experiment from the literature and this study. The experiment and MP2 values were modified from their quoted values in the original papers to ensure the same definition of $E_\textrm{Ov}$ as this paper.  The difference between the reference method and PBE is also shown: $E_\textrm{Ov}^\textrm{Method} -  E_\textrm{Ov}^\textrm{PBE}$, to enable comparison to the quoted Diffusion Monte Carlo (DMC) value. For values of $E_\textrm{Ov}$ or $E_\textrm{Ov}^\textrm{Method} -  E_\textrm{Ov}^\textrm{PBE}$ which are not quoted in their original text, we fill the cell with ``not available'' (n/a). For further consistency with DMC and $\Delta^\textrm{CCSD(T)}_\textrm{PBE}$ studies, $E_\textrm{Ov}^\textrm{PBE}$ has been corrected for its overbinding by using the experimental binding energy of 5.22 eV~\cite{fellerReexaminationAtomizationEnergies1999} (see Sec.\ S14 of the supplementary material).} 
	\begin{ruledtabular}
		\begin{tabular}{lldd}
			Method & Reference & \multicolumn{1}{c}{\textrm{$E^\textrm{Method}_\textrm{Ov}$ (eV)}} & \multicolumn{1}{c}{\textrm{$E_\textrm{Ov}^\textrm{Method} -  E_\textrm{Ov}^\textrm{PBE}$ (eV)}} \\
			\colrule
			Experiment  & Kappers \etal{}\cite{kappersEnsuremathCentersMagnesium1970}       & 9.29 &   \multicolumn{1}{c}{n/a}        \\
			DMC & Ertekin \etal{}\cite{ertekinPointdefectOpticalTransitions2013}         & \multicolumn{1}{c}{n/a}    & 0.5 \\
			$\Delta^\textrm{CCSD(T)}_\textrm{PBE}$ & Richter \etal{}\cite{richterConcentrationVacanciesMetalOxide2013}       & 6.85 & -0.09     \\
			MP2 & Scorza \etal{}\cite{scorzaOxygenVacancySurface1997}         & 7.13 &    \multicolumn{1}{c}{n/a}        \\
			LNO-CCSD(T) & This work & 7.68 & 0.52     
		\end{tabular}
	\end{ruledtabular}
\end{table}

Experimental determination of $E_\textrm{Ov}$ can be challenging due to the many experimental factors that can influence it. For MgO bulk, the single available value of 9.29 eV was obtained from additive coloring experiments by Kappers \etal{}~\cite{kappersEnsuremathCentersMagnesium1970,richterConcentrationVacanciesMetalOxide2013}. These experiments involve heating MgO crystals in Mg vapor under high temperatures and pressures. According to Smakula's formula~\cite{dexterAbsorptionLightAtoms1956}, the measured optical absorption spectra at these temperatures allows for the determination of $E_\textrm{Ov}$ from the relative Mg vapor and Ov concentrations. As seen in Table~\ref{tab:lit_compare}, this value differs by > 1.5 eV from our LNO-CCSD(T) values and other high-level methods. This discrepancy is likely attributed to the uncertainties arising in the experiment. For example, the oscillator strength obtained by Kappers \etal{} differs significantly (> 70 \%) compared to a previous experiment~\cite{chenLuminescenceCenterMgO1969}. Richter \etal{}~\cite{richterConcentrationVacanciesMetalOxide2013} have also attributed this discrepancy to thermal equilibrium not being reached in the experiment. Additionally, some of the assumptions in Smakula's formula can be questionable for ionic solids such as MgO~\cite{dexterAbsorptionLightAtoms1956}. Beyond experimental uncertainties, some of this discrepancy can also arise due to the neglect of temperature effects in the static LNO-CCSD(T) calculations.

Given the above considerations, it is more appropriate to compare high-level references available for the $E_\textrm{Ov}$ in MgO bulk. Ertekin and Grossman~\cite{ertekinPointdefectOpticalTransitions2013} have evaluated $E_\textrm{Ov}$ with DMC for MgO bulk, finding the DMC value to be 0.5 eV higher than PBE. These $E_\textrm{Ov}$ values were computed under Mg-rich conditions, where $E_\textrm{Ov}$ is defined as:
\begin{equation} \label{mg_rich_eov}
E_\textrm{Ov} = E[\textrm{D-MO}] - E[\textrm{P-MO}] +  E[\textrm{MgO}] - E[\textrm{Mg}],
\end{equation}
with $E[\textrm{MgO}]$ and $E[\textrm{Mg}]$ being the total energy of bulk MgO and Mg per formula unit respectively. Compared to the O-rich limit in Eq.~\ref{neweov}, this definition will not suffer from the poor O\textsubscript{2} binding description in $E_\textrm{bind}$. If the O\textsubscript{2} overbinding (0.50 eV/atom) is corrected in the PBE $E_\textrm{Ov}$, our LNO-CCSD(T) value also becomes 0.52 eV higher than PBE, agreeing with DMC values. $\Delta^\textrm{CCSD(T)}_\textrm{PBE}$ embedded cluster calculations by Richter \etal{}~\cite{richterConcentrationVacanciesMetalOxide2013} have obtained an $E_\textrm{Ov}$ value of 6.85 eV, which is 0.83 eV smaller than our LNO-CCSD(T) estimate. 
In Sec.\ S15 of the supplementary material, we show that 0.2 eV of this difference can be attributed to differing lattice parameters and inclusion of structural relaxation around the Ov by performing $\Delta^\textrm{LNO-CCSD(T)}_\textrm{PBE}$ calculations on the same quantum clusters as Richter \etal{}. Remaining differences could arise from the use of differing embedding environments, which our work (Fig.~\ref{fig_04} and Sec.\ S11 A of the supplementary material), alongside others~\cite{chenColorCenterSinglet2020}, have shown cWFT methods to be highly sensitive to, requiring careful convergence. The reference $E_\textrm{Ov}$ value from the $\Delta^\textrm{CCSD(T)}_\textrm{PBE}$ $E_\textrm{Ov}$ study was also found to be smaller than the PBE (with overbinding corrected) value by 0.09 eV, which goes against trends observed from high-level calculations on other metal-oxide systems~\cite{santanaStructuralStabilityDefect2015,santanaDiffusionQuantumMonte2017} as well as the DMC study of Ertekin \etal{}. Explicit MP2 calculations on a converged cluster~\cite{scorzaOxygenVacancySurface1997} give a $E_\textrm{Ov}$ of 7.13 eV, closer to our LNO-CCSD(T) value. The difference w.r.t.\ our LNO-CCSD(T) or LMP2 values arise due to the use of a small (double-zeta quality) basis set; larger basis sets or CBS extrapolations can increase $E_\textrm{Ov}$ significantly by > 0.5 eV w.r.t.\ a double-zeta basis set for MgO bulk (see Sec.\ S4 of the supplementary material).

To our knowledge, the only available high level reference calculation in the TiO\textsubscript{2} system comes from a DFT+GW study~\cite{malashevichFirstprinciplesMathrmDFTGW2014}, with the large number of electrons in Ti making application of methods such as canonical CCSD(T) highly expensive~\cite{chenColorCenterSinglet2020}. Additionally, DMC has not been applied to study the Ov in TiO\textsubscript{2} potentially because the inclusion of a $3d$ transition metal brings additional complications with the starting trial wave-function~\cite{kolorencWaveFunctionsQuantum2010} and the need to validate its pseudopotentials~\cite{krogelPseudopotentialsQuantumMonte2016,annaberdiyevNewGenerationEffective2018}.

\section{Conclusion}

In this work, we discuss a systematic and general approach (named SKZCAM after the authors' initials) for designing small converged quantum clusters for studying the O vacancy with the electrostatic embedded cluster method. When combined with localized orbital correlated wave-function theory methods such as LNO-CCSD(T), this approach allows for accurate determination of the Ov formation energy ($E_\textrm{Ov}$) at a computationally tractable cost. We applied this approach to compute $E_\textrm{Ov}$ values for the bulk and common surface planes of rocksalt MgO and rutile TiO\textsubscript{2} systems. Comparison of these reference values to common DFT XC functionals shows that the studied XC functionals underestimate $E_\textrm{Ov}$ for all studied systems. We observe general improvements in the XC functional errors as Jacob's ladder is ascended, which we find can be correlated to the improvements in the predicted band gaps of the XC functionals. %
Of the XC functionals studied, the double-hybrid B2PLYP functional gives the best performance, with a mean absolute error within the error bars of the reference calculation. 
Other BLYP-based functionals, such as BLYP and B3LYP are also found to perform well within their respective Jacob's ladder rungs, alongside the meta-GGA SCAN and the hybrid $\omega$B97X functionals. 

Although this work focuses on the Ov formation energy, the simple and intuitive nature of the outlined protocol should allow its application to other chemical problems, including molecular adsorption, spectroscopic quantities or complex (ternary) metal-oxide systems. Furthermore, the approach can be automated, making it amenable for integration into existing high-throughput calculation frameworks. Such high-throughput frameworks~\cite{kumagaiInsightsOxygenVacancies2021} for point defects or molecular adsorbates are highly desirable. For example, it can be used to screen for target applications in catalysis~\cite{hinumaDensityFunctionalTheory2018,demlIntrinsicMaterialProperties2015} or to produce large reference databases to validate current XC functionals~\cite{rezacBenchmarkCalculationsInteraction2016}. Beyond the applications, we have defined a rigorous and well-founded framework for controlling the shape of the embedded quantum clusters. This framework can be valuable in not only defining the quantum cluster in electrostatic embedded cluster approaches, but also in other embedding approaches~\cite{leeProjectionBasedWavefunctioninDFTEmbedding2019,schaferLocalEmbeddingCoupled2021,yuImplementationDensityFunctional2015,hegelyExactDensityFunctional2016,bernsteinHybridAtomisticSimulation2009}, where multiple clusters, corresponding to different levels of theory, may have to be defined simultaneously.

\section*{Supplementary material}
See the supplementary material for a detailed compilation of the obtained results as well as further data and analysis to support the points made throughout the text.

\begin{acknowledgments}
We are grateful for: resources provided by the Cambridge Service for Data Driven Discovery (CSD3) operated by the University of Cambridge Research Computing Service (\href{www.csd3.cam.ac.uk}{www.csd3.cam.ac.uk}), provided by Dell EMC and Intel using Tier-2 funding from the Engineering and Physical Sciences Research Council (capital grant EP/P020259/1), and DiRAC funding from the Science and Technology Facilities Council (\href{www.dirac.ac.uk}{www.dirac.ac.uk}); computational resources granted by the UCL Myriad and Kathleen High Performance Computing Facility (Myriad@UCL and Kathleen@UCL), and associated support service; computational support from the UK Materials and Molecular Modelling Hub, which is partially funded by EPSRC (EP/P020194 and EP/T022213); and computational support from the UK national high performance computing service, ARCHER 2. Both the UK Materials and Molecular Modelling Hub and ARCHER 2 access was obtained via the UKCP consortium and funded by EPSRC grant ref EP/P022561/1.
\end{acknowledgments}

\section*{Data availability}
The data that supports the findings of this study are available within the article and its supplementary material. The input and output files associated with this study and all analysis can be found on GitHub at \href{https://github.com/benshi97/Data_Embedded_Cluster_Protocol}{benshi97/Data\_Embedded\_Cluster\_Protocol} and \href{https://mybinder.org/v2/gh/benshi97/Data_Embedded_Cluster_Protocol/HEAD?labpath=analyse_data.ipynb}{Binder}. The scripts used to generate the quantum clusters with the SKZCAM approach is also provided in a separate GitHub repository at \href{https://github.com/benshi97/SKZCAM-Protocol}{benshi97/SKZCAM-Protocol}  and \href{https://mybinder.org/v2/gh/benshi97/SKZCAM-Protocol/HEAD?labpath=generate_clusters.ipynb}{Binder}.

\end{document}

\title{Supplementary material: General embedded cluster protocol for accurate modeling of oxygen vacancies in metal-oxides}

\author{Benjamin X. Shi}
\affiliation{Yusuf Hamied Department of Chemistry, University of Cambridge, Lensfield Road, Cambridge CB2 1EW, United Kingdom}%

\author{Venkat Kapil}
\affiliation{Yusuf Hamied Department of Chemistry, University of Cambridge, Lensfield Road, Cambridge CB2 1EW, United Kingdom}%
\affiliation{Churchill College, University of Cambridge, Storey's Way, Cambridge CB3 0DS}%

\author{Andrea Zen}
\affiliation{Dipartimento di Fisica Ettore Pancini, Universit\`{a} di Napoli Federico II, Monte S. Angelo, I-80126 Napoli, Italy}
\affiliation{Department of Earth Sciences, University College London, Gower Street, London WC1E 6BT, United Kingdom}

\author{Ji Chen}
\affiliation{School of Physics, Peking University, Beijing, 100871, China}

\author{Ali Alavi}
\affiliation{Max Planck Institute for Solid State Research, Heisenbergstra{\ss}e 1, 70569 Stuttgart, Germany}
\affiliation{Yusuf Hamied Department of Chemistry, University of Cambridge, Lensfield Road, Cambridge CB2 1EW, United Kingdom}%

\author{Angelos Michaelides}
\affiliation{Yusuf Hamied Department of Chemistry, University of Cambridge, Lensfield Road, Cambridge CB2 1EW, United Kingdom}%
\affiliation{Department of Physics and Astronomy, University College London, Gower Street, London, WC1E 6BT, United Kingdom}
\affiliation{Thomas Young Centre and London Centre for Nanotechnology, 17-19 Gordon Street, London WC1H 0AH, United Kingdom}

\maketitle

\tableofcontents
\newpage

\section{XC functional dependence of O\textsubscript{2} binding energies}

The O\textsubscript{2} binding energies quoted in Table \ref{tab:o2_bind} were calculated using:
\begin{equation}
E_\textrm{bind} = 2E[\textrm{O}] -  E[\textrm{O\textsubscript{2}}],
\end{equation}
where $E[\textrm{O}]$ and $E[\textrm{O\textsubscript{2}}]$ are the total energies of the O atom and O\textsubscript{2} molecule, both in the unrestricted triplet state. We have used the same O\textsubscript{2} geometry obtained from R2SCAN structural optimizations, with the O atoms separated by 1.207 \AA{}.

\begin{table}[h]
	\setlength\extrarowheight{-8pt}
	\caption{\label{tab:o2_bind} O\textsubscript{2} binding energy predicted by DFT XC functionals as well as at the B2PLYP, LMP2, LNO-CCSD and LNO-CCSD(T) correlated wave-function theory (cWFT) levels. DFT calculations were performed at the def2-QZVPP level. cWFT methods (including B2PLYP) were computed through a CBS(TZVPP/QZVPP) extrapolation with a 1s frozen core on the O atoms.}
	\begin{ruledtabular}
		\begin{tabular}{lc}
			XC functional & $E_\textrm{bind}$ (eV) \\ \midrule
			PBE           & 6.22       \\
			BP86          & 6.15       \\
			BLYP          & 5.86       \\
			TPSS          & 5.49       \\
			SCAN          & 5.52       \\
			PBE0          & 5.38       \\
			HSE06         & 5.33       \\
			B3LYP         & 5.32       \\
			$\omega$B97X         & 5.41       \\
			B2PLYP       & 5.36       \\
			LMP2          & 5.65       \\
			LNO-CCSD      & 4.80       \\
			LNO-CCSD(T)   & 5.16       \\
			Experiment~\cite{fellerReexaminationAtomizationEnergies1999}          & 5.22      
		\end{tabular}
	\end{ruledtabular}
\end{table}

\FloatBarrier

\section{XC functional dependence of the O vacancy (Ov) relaxation energy in rutile TiO\textsubscript{2}}
Table \ref{tab:relax_ene} shows the change in relaxation energy for different semilocal XC functionals in TiO\textsubscript{2} bulk and its $(110)$ surface. The relaxation energy, $E_\textrm{rel.}$, is defined as:
\begin{equation}
E_\textrm{rel.} = E[\textrm{D-MO}] - E[\textrm{r-D-MO}],
\end{equation}
where $E[\textrm{D-MO}]$ and $E[\textrm{r-D-MO}]$ are the energies of the \textit{unrelaxed} and \textit{relaxed} defected structures, containing an Ov, respectively. We observe a wide range of around 0.45 and 0.22 eV in the relaxation energies for the bulk and surface just at the semilocal XC functional level.

\begin{table}[h]
	\caption{\label{tab:relax_ene} Relaxation energy of the Ov defect in the closed-shell singlet state in TiO\textsubscript{2} bulk and its $(110)$ surface. The bulk calculations were performed in a 192 atom $(2\sqrt{2} \times 2 \sqrt{2} \times 4)$ supercell whilst the surface calculations were performed in a $p(2\times4)$ asymmetric supercell slab where the top two layers were allowed to relax.}
	\begin{ruledtabular}
		\begin{tabular}{lcc}
			XC functional& TiO\textsubscript{2} bulk & TiO\textsubscript{2} surface \\ \midrule
			LDA    & 1.29      & 2.29         \\
			PBE    & 1.45      & 2.27         \\
			PBEsol & 1.32      & 2.30         \\
			R2SCAN & 1.00      & 2.08        
		\end{tabular}
	\end{ruledtabular}
\end{table}

\FloatBarrier

\section{Convergence of point charge and effective core potential boundary length}

Table \ref{tab:ecppclength} quotes the change in $E_\textrm{Ov}$ for a embedded cluster of the TiO\textsubscript{2} surface as the point charge (PC) and effective core potential (ECP) regions are changed in length. TiO\textsubscript{2} surface is used as the example as it represents the system that is most difficult to converge.  We see that the changes in $E_\textrm{Ov}$ are all small ($\sim$0.01 eV) for substantial changes in both the PC and ECP regions, suggesting our chosen parameters of 7 \AA{} and 40 \AA{} are sufficient for this system.
\begin{table}[h]
	\caption{\label{tab:ecppclength} Change in O vacancy formation energy ($E_\textrm{Ov}$) as point charge (PC) and effective core potential (ECP) regions of an embedded cluster are changed. Values are computed at the PBE-DFT level with a def2-SVP basis set. The quantum cluster region consists of 31 Ti ions, taken from the quantum cluster series produced via the SKZCAM approach.} 
	\begin{ruledtabular}
		\begin{tabular}{dcd}
			\multicolumn{1}{c}{\textrm{ECP (\AA{})}} & PC (\AA{}) & \multicolumn{1}{c}{\textrm{$E_\textrm{Ov}$}}  \\ \midrule
			7.0          & 30        & 5.501		 \\
			7.0          & 40        & 5.513		 \\
			10.6       & 40        & 5.513		
		\end{tabular}
	\end{ruledtabular}
\end{table}

\FloatBarrier

\section{Basis set convergence}
\subsection{DFT}
Our basis set tests on the smallest converged clusters (SCCs) of the four studied systems in Table \ref{tab:dftconvergence} shows that at the PBE0 level, def2-TZVPP and def2-QZVPP $E_\textrm{Ov}$ values are within 0.02 eV, indicating that they are both converged. def2-SVP shows large differences up to 0.3 eV from the def2-QZVPP basis set.

\begin{table}[h]
	\caption{\label{tab:dftconvergence} Convergence of the O vacancy formation energy ($E_\textrm{Ov}$) with basis set size in the SCCs of rocksalt MgO and rutile TiO\textsubscript{2}, in their bulk and surface, for the def2-SVP, def2-TZVPP and def2-QZVPP basis sets. Calculations were performed at the PBE0 level.}
	\begin{ruledtabular}
		\begin{tabular}{lcccc}
			& MgO bulk & MgO surface & TiO\textsubscript{2} bulk & TiO\textsubscript{2} surface \\ \midrule
			def2-SVP   & 7.33     & 6.66        & 6.10      & 5.39         \\
			def2-TZVPP & 6.97     & 6.37        & 6.06      & 5.42         \\
			def2-QZVPP & 6.99     & 6.39        & 6.07      & 5.41        
		\end{tabular}
	\end{ruledtabular}
\end{table}
\FloatBarrier

\newpage
\subsection{LNO-CCSD(T)}

\begin{figure*}[h]
	\includegraphics{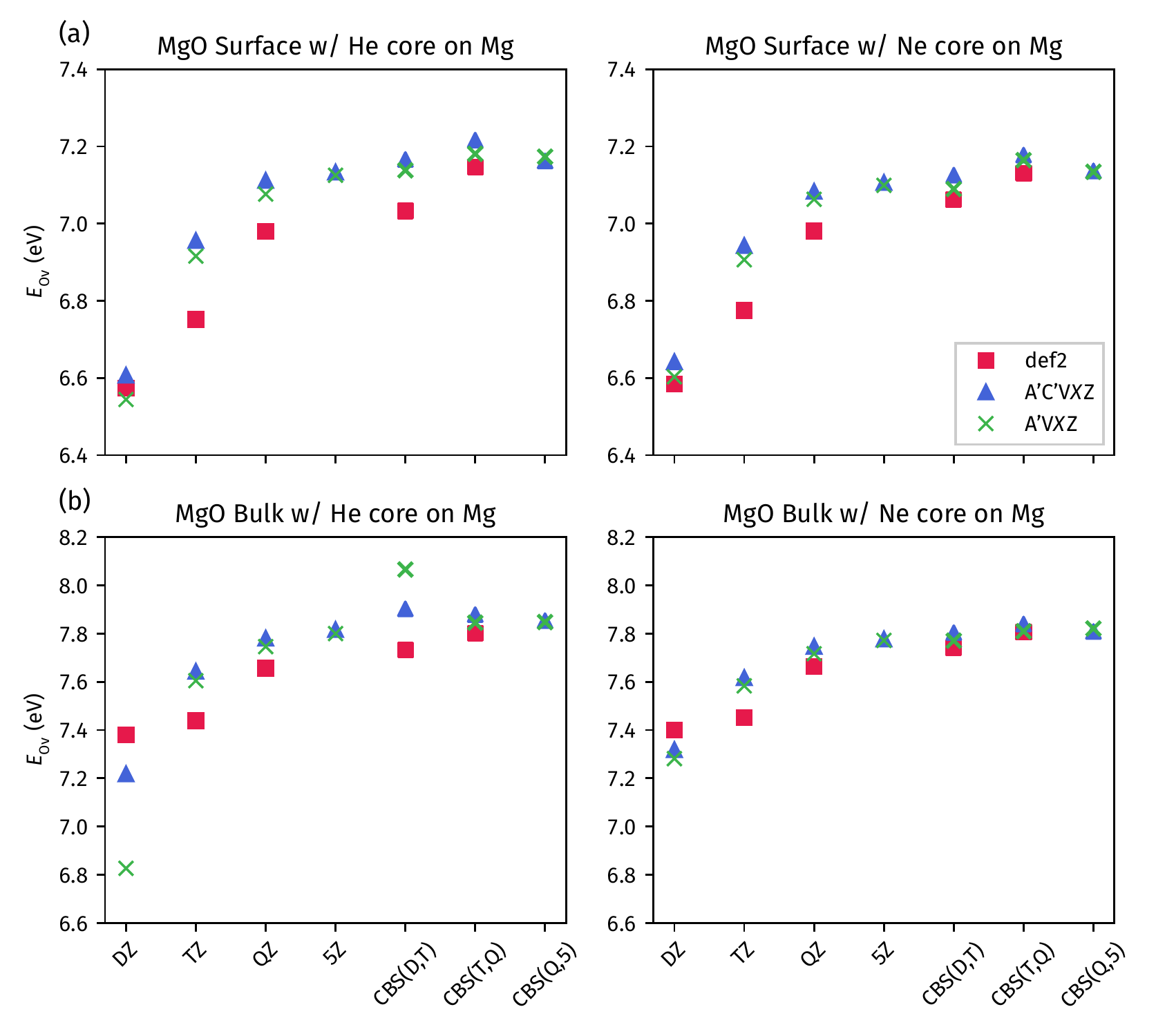}%
	\caption{\label{fig:mgo_basis_conv} Convergence of the O vacancy formation energy ($E_\textrm{Ov}$) with basis set size in (a) bulk and (b) surface MgO for the def2, A'V$X$Z and A'C'V$X$Z basis set families. A He and Ne frozen core on Mg is tested for both sets. Two point basis set extrapolations for DZ/TZ, TZ/QZ and QZ/5Z combos are also plotted for the respective basis set families. These calculations were all performed with ``Tight'' LNO thresholds.}
\end{figure*}

\begin{figure*}[h]
	\includegraphics{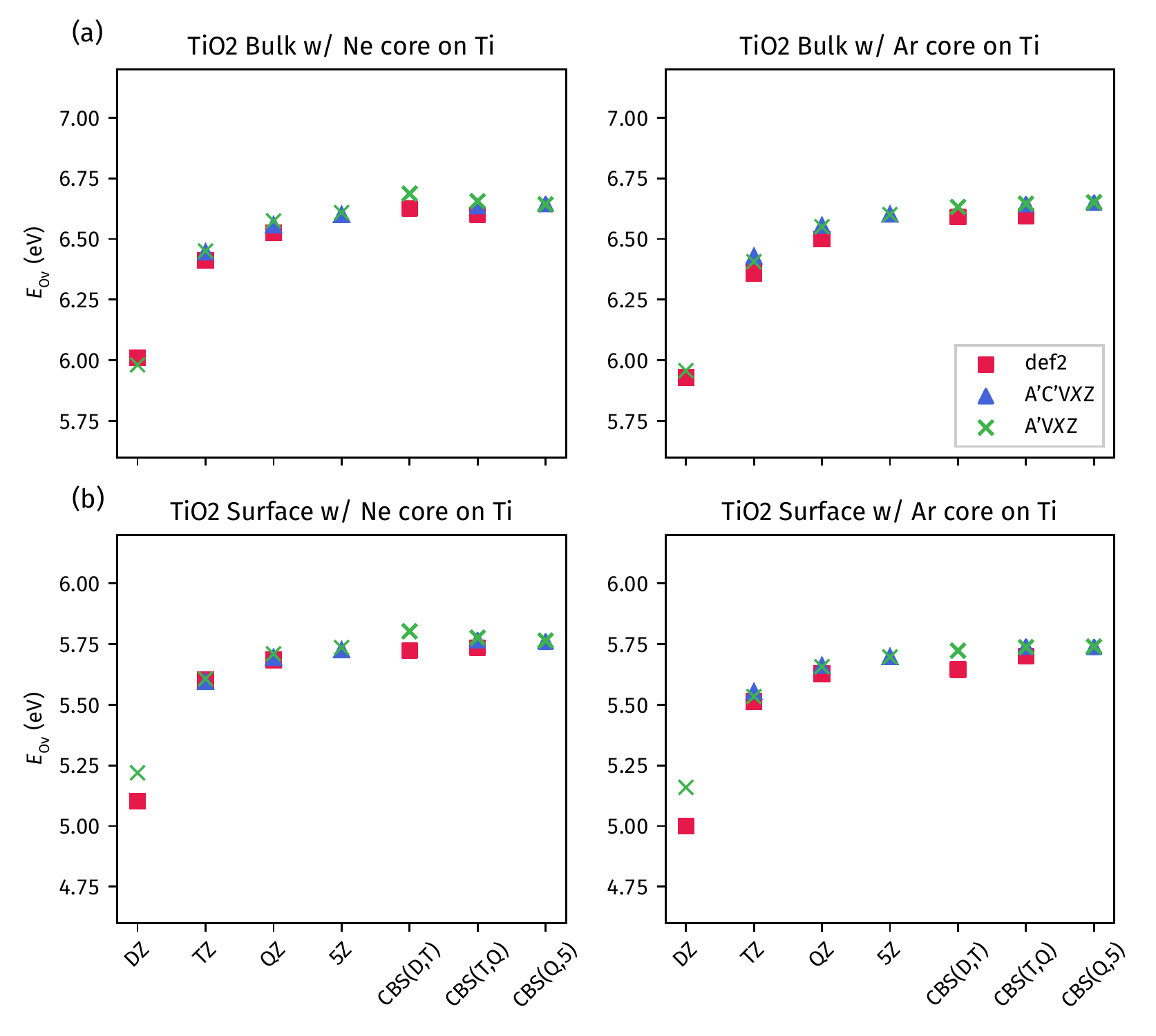}%
	\caption{\label{fig:tio2_basis_conv} Convergence of the O vacancy formation energy ($E_\textrm{Ov}$) with basis set size in (a) bulk and (b) surface TiO\textsubscript{2} for the def2, A'V$X$Z and A'C'V$X$Z basis set families. A Ne and Ar frozen core on Ti is tested for both sets. Two point basis set extrapolations for DZ/TZ, TZ/QZ and QZ/5Z combos are also plotted for the respective basis set families. These calculations were all performed with ``Tight'' LNO thresholds.}
\end{figure*}

During our basis set tests, we tested three commonly used basis sets: def2, aug'-cc-pV$X$Z (A'V$X$Z) and aug'-cc-pwC'V$X$Z (A'C'V$X$Z) basis set families. For the latter two basis sets, only the O ions have been augmented with diffuse functions, with the cc-pV$X$Z and cc-pwCV$X$Z basis sets placed on the metal cation for the A'V$X$Z and A'C'V$X$Z basis sets respectively. In Figure \ref{fig:mgo_basis_conv}, the basis set convergence of these three basis set families at the LNO-CCSD(T) level are plotted for a MgO bulk (with 6 Mg ions) and surface (with 5 Mg ions) quantum cluster -- small quantum clusters obtained through the SKZCAM approach. We have also compared $E_\textrm{Ov}$ for small He or large Ne frozen core on the Mg cation. As shown in these tests, the use of the CBS(TZVPP/QZVPP) basis set extrapolation with the def2 basis set with a large Ne frozen core shows excellent agreement (to less than 0.1 eV) compared to the most accurate calculation: CBS(QZ/5Z) with the A'C'V$X$Z basis set family using a small frozen core. We have performed a similar comparison for the TiO\textsubscript{2} systems in Fig.~\ref{fig:tio2_basis_conv} and obtain good agreement of around 0.1 eV as well. The quantum clusters used for the bulk and surface consisted of 3 and 2 Ti ions respectively, all obtained from the SKZCAM approach.

\FloatBarrier

\FloatBarrier

\newpage

\section{Basis set and frozen core correction from small clusters}
We computed the change in $E_\textrm{Ov}$ at the LNO-CCSD(T) level moving from CBS(TZVPP/QZVPP) with large frozen core (Ar for Ti and Ne for Mg) to CBS(A'C'VTZ/A'C'VQZ) with small frozen core (Ne for Ti and He for Mg) for several small tractable quantum clusters, created through the SKZCAM approach, in the bulk and common surface planes of rocksalt MgO and rutile TiO\textsubscript{2}. These clusters were used to estimate any deficiencies for using the CBS(TZVPP/QZVPP) with large frozen core in the SCCs of the four systems, which we use to compute a correction. For MgO surface, we have been able to compute the CBS(A'C'VTZ/A'C'VQZ) value for the explicit SCC and this value was quoted in our final best estimates. For the other clusters, we have taken the average difference from the computed clusters as a correction to CBS(TZVPP/QZVPP) values on the SCC. 

\begin{table}[h]
	\caption{\label{tab:correction_size} $E_\textrm{Ov}$ difference (in eV) at the LNO-CCSD(T) level moving from CBS(TZVPP/QZVPP) with large frozen core (Ar for Ti and Ne for Mg) to CBS(A'C'VTZ/A'C'VQZ) with small frozen core (Ne for Ti and He for Mg) for small MgO and TiO\textsubscript{2} quantum clusters. For some clusters, such as those consisting of the first RDF shell in MgO surface and TiO\textsubscript{2} bulk, we have not computed this difference because these structures are ``unphysical'' since the O vacancy is not fully coordinated by metal cations. We have also computed the corrections at the B2PLYP, LMP2 and LNO-CCSD levels.}
	\begin{ruledtabular}
		\begin{tabular}{lcccc}
			\# of RDF shells   & MgO bulk & MgO surface & TiO\textsubscript{2} bulk & TiO\textsubscript{2} surface \\ \midrule
			1                 & 0.06     &             &           & 0.08         \\
			2                 &          & 0.09        & 0.07      & 0.07         \\
			3                 &          & 0.15        & 0.10      &              \\
			4                 &          & 0.16        &           &              \\ \midrule
			Final correction - LNO-CCSD(T) & 0.06     & 0.16        & 0.09      & 0.07 \\
			Final correction - LNO-CCSD &  0.06 &         0.14    &   0.07 &          0.14 \\
		Final correction - LMP2    &         0.04    &     0.13     & -0.14     &    -0.21 \\
		Final correction - B2PLYP  &         0.00    &     0.05     &  -0.02      &   -0.04
		\end{tabular}
	\end{ruledtabular}
\end{table}

\FloatBarrier

\newpage

\section{XC functional dependence of rutile TiO\textsubscript{2} and rocksalt MgO lattice parameters}
\begin{table}[h]
	\caption{Lattice parameters for the conventional unit cells of rutile TiO\textsubscript{2} ($a$, $c$ and $u$) and rocksalt MgO ($a$) predicted by LDA, PBE, R2SCAN and HSE06 exchange-correlation functionals in DFT compared to experiment. Calculations were performed in VASP with an 800 eV energy cutoff with $9 \times 9 \times 14$ and $9\times 9 \times 9$ $k$-point meshes for TiO\textsubscript{2} and MgO respectively. R2SCAN gives the best agreement of the lattice parameter out of all the studied functionals. }
	\begin{ruledtabular}	
		\begin{tabular}{l|cccc|cc}
			\multirow{2}{*}{Functional}& \multicolumn{4}{c| }{Rutile TiO\textsubscript{2}}     & \multicolumn{2}{c}{Rocksalt MgO} \\ \cline{2-7}  
			& a      & c      & u      & MAE (\%) & a            & Error (\%)        \\ \midrule
			LDA              & 4.553 & 2.922 & 0.3038 & 0.70     & 4.163        & 1.25              \\
			PBE              & 4.646 & 2.967 & 0.3050 & 0.61     & 4.251        & 0.83              \\
			R2SCAN           & 4.600 & 2.960 & 0.3045 & 0.18     & 4.206        & 0.23              \\
			HSE06            & 4.583 & 2.945 & 0.3052 & 0.18     & 4.203        & 0.31              \\ \midrule
			Experiment~\cite{burdettStructuralelectronicRelationshipsInorganic1987,madelungSemiconductorsDataHandbook2012} & 4.587 & 2.954 & 0.3047 &          & 4.216        &                  
		\end{tabular}
	\end{ruledtabular}
\end{table}

\FloatBarrier

\section{Bulk limit Ov formation energies from supercell calculations}

Table \ref{eovbulklim} lists the bulk limit values used in Fig.\ 3 of the main text. Other than TiO\textsubscript{2} surface, the other systems only required PBE bulk limit values. PBE0 calculations were performed for TiO\textsubscript{2} surface for reasons discussed in Sec.\ III B of the main text. These calculations are expensive and to circumvent some of these computational costs, a smaller $(2 \times 4)$ supercell slab was used for the PBE0 calculations with finite size correction to the $(2 \times 6)$ supercell slab approximated by PBE. The difference of these two numbers for PBE is added as a correction to the $(2 \times 4)$ supercell calculation of PBE0 to approximate its results at the bulk limit.

\begin{table}[h]
	\caption{\label{eovbulklim} O vacancy formation energy ($E_\textrm{Ov}$) computed for rocksalt MgO and rutile TiO\textsubscript{2}, for both their bulk and surface forms. The $E_\textrm{Ov}$ values were obtained using the corresponding DFT O\textsubscript{2} molecular binding energy from Table~\ref{tab:o2_bind}.}
	\begin{ruledtabular}
		\begin{tabular}{lccccc}
			& MgO bulk & MgO surface & TiO\textsubscript{2} bulk & TiO\textsubscript{2} surface $(2\times4)$ & TiO\textsubscript{2} surface $(2\times6)$ \\ \midrule
			PBE  & 6.64     & 6.12        & 5.61      & 5.10               & 5.07               \\
			PBE0 &          &             &           & 5.47               & 5.44              
		\end{tabular}
	\end{ruledtabular}
\end{table}

\FloatBarrier

\section{Complexity of designing a quantum cluster series}

\subsection{Range of shapes for a given size}
\begin{figure*}[h]
	\includegraphics{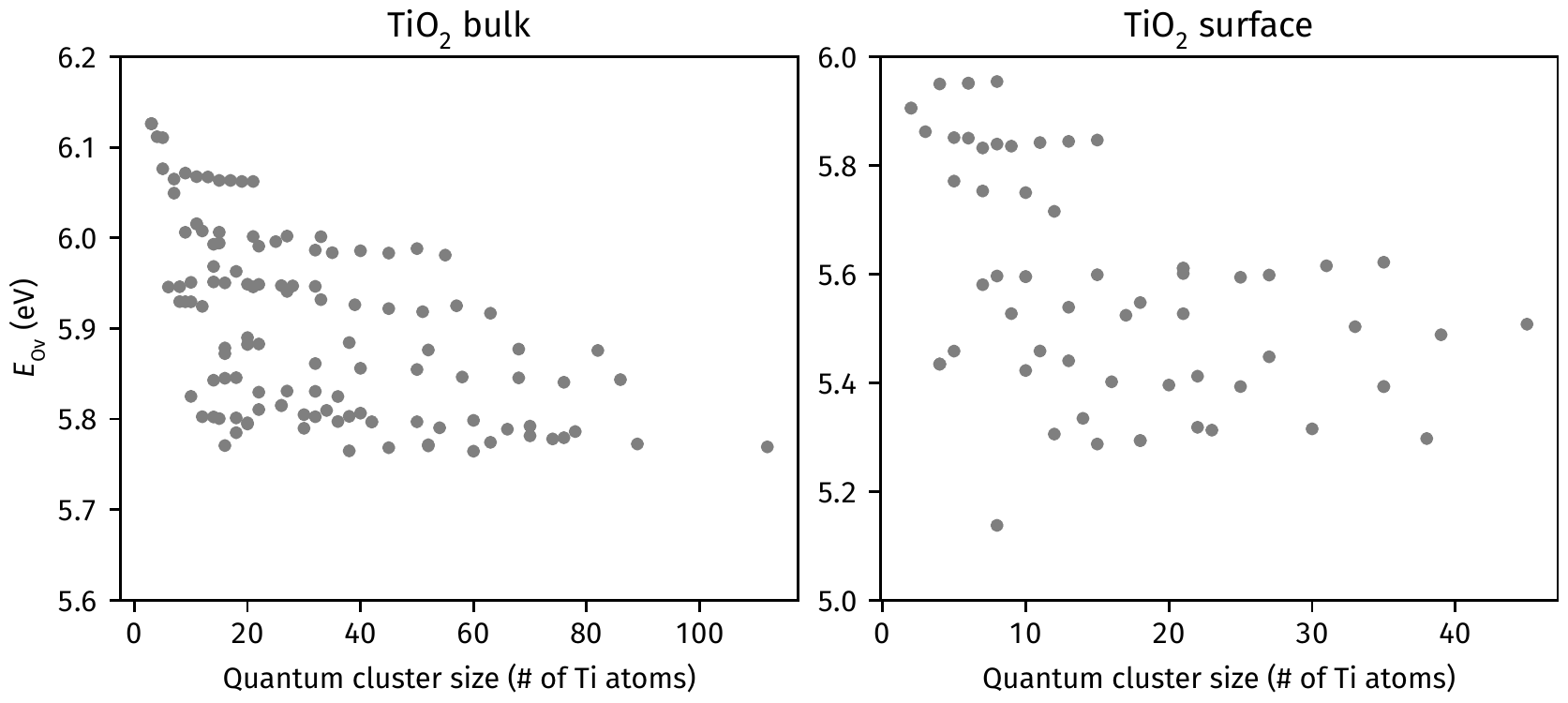}
	\caption{\label{fig:range_EOv} Illustrating the range of O vacancy formation energy ($E_\textrm{Ov}$) that can be found for different shapes for a given size of quantum cluster for both rutile TiO\textsubscript{2} bulk and surface systems.}
\end{figure*}

In Fig.\ \ref{fig:range_EOv}, the O vacancy formation energy for randomly shaped quantum clusters of rutile TiO\textsubscript{2}, both in its bulk and common surface plane is plotted. All of the studied quantum clusters are negatively charged with O anions placed to ensure no dangling bonds on the Ti cations, with the only difference being different spatial arrangements of the Ti cations. These $E_\textrm{Ov}$ are computed at the def2-SVP PBE-DFT level and we find that there can be a wide range of over 0.3 and 0.8 eV for the bulk and surface respectively.

\FloatBarrier

\subsection{Ambiguity in the O anion spatial arrangement for stoichiometric clusters}
\begin{figure*}[h]
	\includegraphics{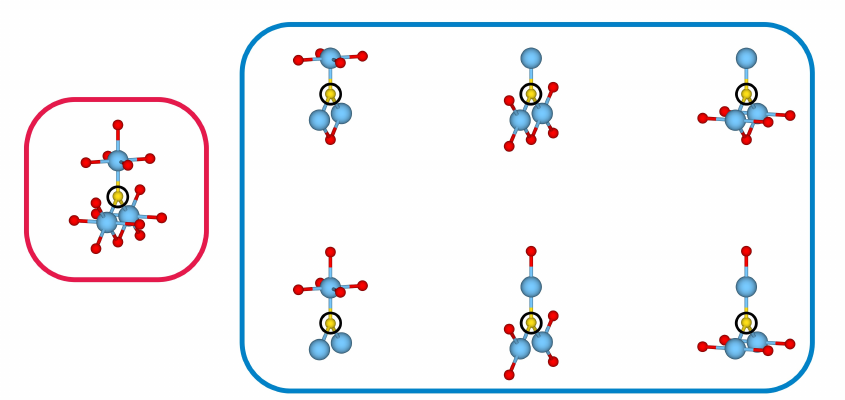}%
	\caption{\label{fig:O_stoic} The quantum cluster, created through the SKZCAM approach, involving 3 Ti ions for bulk rutile TiO\textsubscript{2} is visualized in the red box. In the blue box, we give six possible stoichiometric clusters involving the same 3 Ti ions, but with different O anion spatial arrangements.}
\end{figure*}

For the 3 Ti ion quantum cluster in Fig.\ 2 of the main text, we have computed $E_\textrm{Ov}$ for several quantum clusters. There is the singular red point corresponding to the cluster created by the SKZCAM approach. There is no ambiguity in the O anion positions since we have used the robust rubric of removing all dangling bonds on the Ti cations. On the other hand, for a stoichiometric quantum cluster of the same size, there can be several O anion configurations for the same Ti cations and we show six such possibilities in Fig.~\ref{fig:O_stoic}. These clusters have a wide range of $E_\textrm{Ov}$ spanning over 1.0 eV in Fig.\ 2 (c) of the main text, with no intuition and easy way to find the ``optimal'' O anion configuration. The number of possible configurations is expected to become even larger as we go to larger quantum clusters.

\FloatBarrier

\section{Finding smaller converged clusters with the SKZCAM approach}

\begin{figure}[h]
	\includegraphics{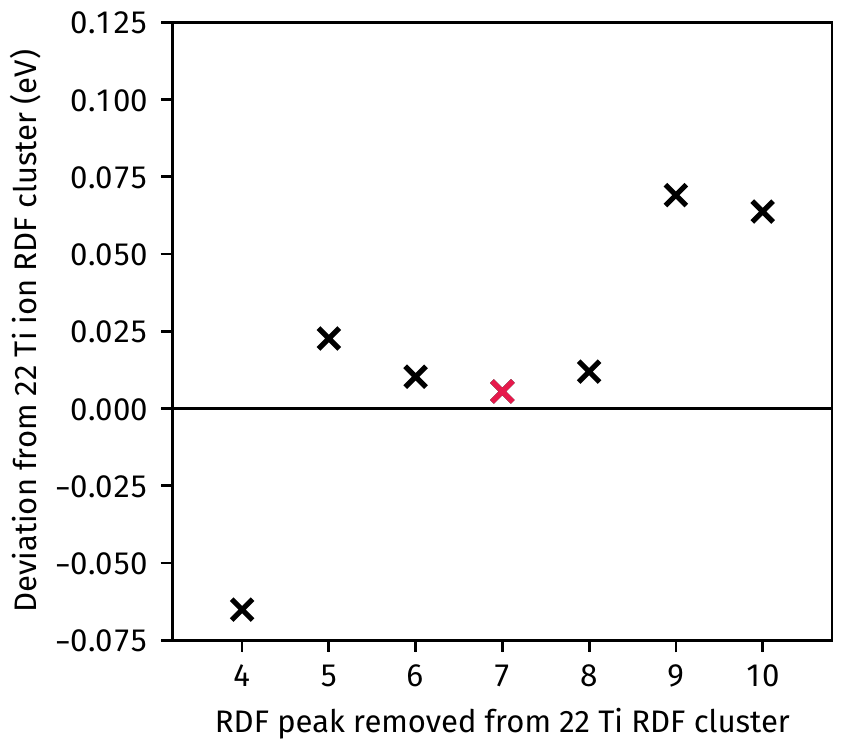}%
	\caption{\label{fig_s1} Change in O vacancy formation energy ($E_\textrm{Ov}$) when Ti cation shells corresponding to different RDF peaks are removed from the 22 Ti ion cluster.}
\end{figure}

In Sec.\ III A of the main text, we have described the SKZCAM approach to find small converged quantum clusters. Based on the radial parameter (i.e.\ including RDF shells based on distance from the Ov), we find a 22 Ti ion quantum cluster, consisting of the first ten RDF shells. From this quantum cluster, smaller converged clusters can be found by considering the removal of RDF shells closer to the Ov. In Fig.~\ref{fig_s1}, we show the change in $E_\textrm{Ov}$ when shells, numbered by distance from Ov, are removed from the quantum cluster. We find that removal of the 7\textsuperscript{th} RDF shell leads to little change in $E_\textrm{Ov}$ at a substantial decrease in size to a cluster with 18 Ti ions -- the smallest converged cluster described in the main text.

\FloatBarrier

\section{Basis set correction for RDF size convergence in MgO}
To aid the speed of calculations and because larger basis sets suffer from linear dependencies at larger cluster sizes, we have performed size convergence calculations in MgO using the def2-SVP basis set. To allow comparison to the bulk limit, we have shifted the def2-SVP $E_\textrm{Ov}$ values for all of the studied quantum clusters by a correction to the def2-TZVPP level computed from their average difference for a few clusters beyond the SCC, as shown in Table \ref{tab:svp2tzvppcorr}.

\begin{table}[h]
	\caption{\label{tab:svp2tzvppcorr} The difference in the def2-SVP and def2-TZVPP O vacancy formation energy ($E_\textrm{Ov}$) for a few quantum clusters larger than the smallest converged cluster for MgO bulk and surface. A correction is computed from the average difference, which we then apply to the def2-SVP $E_\textrm{Ov}$ of all the studied clusters for the respective system in Fig.\ 3 of the main text.}
	\begin{ruledtabular}
		\begin{tabular}{lcccc}
			\multicolumn{5}{c}{MgO   bulk}                                                                                                   \\ \midrule
			\# of Mg ions & SVP                       & TZVPP                     & Difference                 & Correction              \\ \midrule
			38                & 7.010                     & 6.637                     & -0.373                     & \multirow{2}{*}{-0.372} \\
			68                & 7.009                     & 6.638                     & -0.371                     &                         \\ \midrule
			\multicolumn{5}{c}{MgO surface}                                                                                                  \\ \midrule
			\# of Mg ions & SVP                       & TZVPP                     & Difference                 & Correction              \\ \midrule
			17                & 6.380                     & 6.121                     & -0.258                     & \multirow{3}{*}{-0.255} \\ 
			21                & 6.369 & 6.117 & -0.253     &                         \\
			25                & 6.364 & 6.112 & -0.253     &                               
		\end{tabular}
	\end{ruledtabular}
\end{table}

\FloatBarrier

\section{Quantum cluster RDF size convergence}
\subsection{MgO}
Tables \ref{mgo_bulk_size} and \ref{mgo_surf_size} shows the change in $E_\textrm{Ov}$ at the LMP2 level as the quantum cluster size is increased for MgO bulk and surface respectively. For both of these systems, the smallest converged cluster (SCC) obtained at the PBE-DFT level, highlighted in bold, has errors w.r.t.\ the largest studied cluster of less than 0.05 eV. LNO-CCSD(T) results are also shown for the MgO surface in Table \ref{mgo_surf_size} and we observe a similar size convergence behavior as at the LMP2 level.

\begin{table}[h]
	\caption{\label{mgo_bulk_size} Change in O vacancy formation energy ($E_\textrm{Ov}$) with cluster size -- generated using the SKZCAM approach -- at the LMP2 level for bulk MgO. Errors were calculated w.r.t.\ the largest computationally tractable cluster. Calculations were performed in MRCC 2020 at the def2-TZVPP level with ``Normal'' LNO threshold settings.}
	\begin{ruledtabular}
		\begin{tabular}{cccc}
			\# of RDF peaks & \# of Mg ions & LMP2  & \multicolumn{1}{c}{\textrm{Error}} \\ \midrule
			1               & 6             & 7.68 & 0.38                           \\
			2               & 14            & 7.48 & 0.17                           \\
			\textbf{3}               & \textbf{38}            & $\mathbf{7.33}$  & $\mathbf{0.02}$                         \\ \midrule
			4               & 68            & 7.31 & 0.00                           \\
		\end{tabular}
	\end{ruledtabular}
\end{table}

\begin{table}[h]
	\caption{\label{mgo_surf_size} Change in O vacancy formation energy ($E_\textrm{Ov}$) with cluster size -- generated using the SKZCAM approach -- at the LMP2 and LNO-CCSD(T) level for the $(001)$ MgO surface. Calculations were performed in MRCC 2020 at the def2-TZVPP level with ``Normal'' LNO threshold settings.}
	\begin{ruledtabular}
		\begin{tabular}{cccrcr}
			\# of RDF peaks & \# of Mg ions & LMP2          & Error         & LNO-CCSD(T)   & Error         \\ \midrule
			1               & 4             & 7.14          & 0.25          & 6.84          & 0.26          \\
			2               & 5             & 7.05          & 0.17          & 6.76          & 0.18          \\
			3               & 9             & 6.98          & 0.09          & 6.68          & 0.10          \\
			\textbf{4}      & \textbf{17}   & \textbf{6.93} & \textbf{0.05} & \textbf{6.62} & \textbf{0.04} \\
			5               & 21            & 6.88          & -0.01         & 6.55          & -0.03         \\
			6               & 25            & 6.85          & -0.04         & 6.53          & -0.05         \\
			7               & 29            & 6.84          & -0.04         & 6.52          & -0.06         \\
			8               & 33            & 6.87          & -0.01         & 6.56          & -0.02         \\
			9               & 41            & 6.88          & -0.01         & 6.57          & -0.01         \\ \midrule
			10              & 42            & 6.88          & 0.00          & 6.58          & 0.00          \\ 
		\end{tabular}
	\end{ruledtabular}
\end{table}

\FloatBarrier

\newpage
\subsection{TiO\textsubscript{2}}

\begin{figure}[h]
	\includegraphics{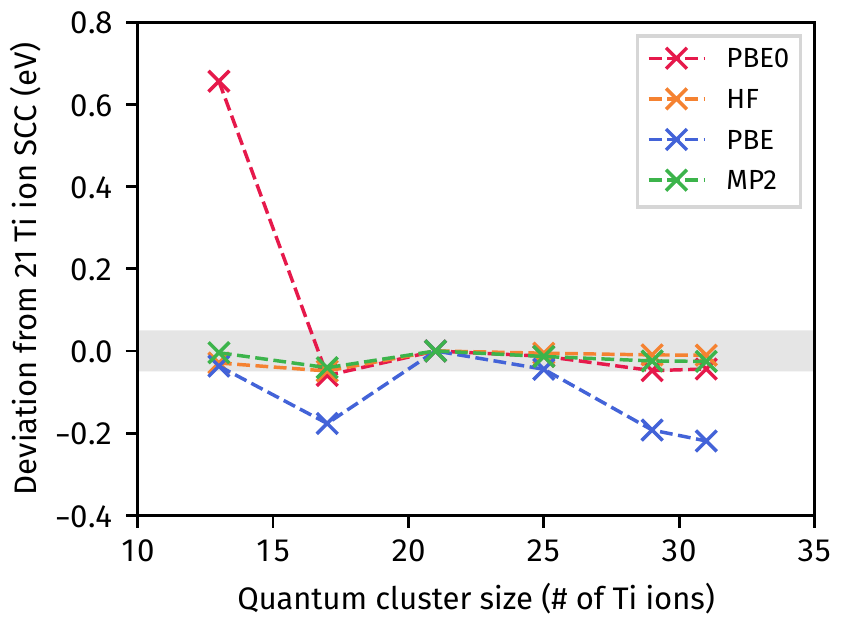}%
	\caption{\label{fig:tio2_surf_size} Deviation in the O vacancy formation energy ($E_\textrm{Ov}$) from the 21 Ti ion SCC for quantum clusters larger and smaller than the SCC of TiO\textsubscript{2} surface at the PBE, PBE0, HF and LMP2 levels of theory.}
\end{figure}

Figure \ref{fig:tio2_surf_size} plots the deviation of $E_\textrm{Ov}$ w.r.t.\ the 21 Ti ion SCC for TiO\textsubscript{2} surface. For clusters larger than the SCC, there is little change in the HF, LMP2 and PBE0 levels, all to within the gray 0.05 eV error bar. PBE shows strong deviations due to the aforementioned odd-even oscillations in Sec.~\ref{sec:odd-even}. At the SCC size, the PBE $E_\textrm{Ov}$ of 5.00 eV shows good agreement to the supercell calculation (5.07 eV) and we have established good agreement (to within 0.03 eV) for the PBE0 functional in the main text.

\FloatBarrier

\section{\label{sec:odd-even} Odd-even oscillations in TiO\textsubscript{2} surface}

\begin{figure}[h]
	\includegraphics{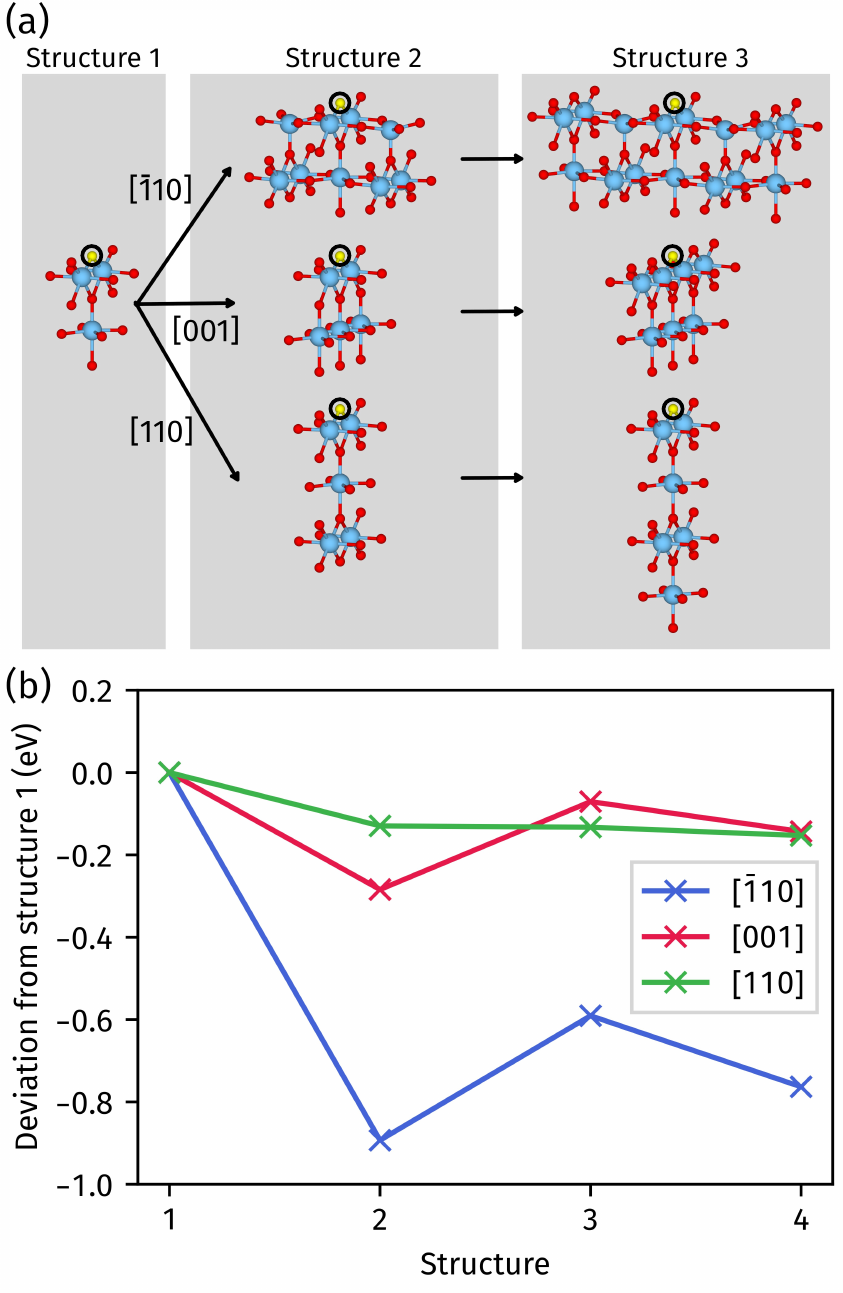}%
	\caption{\label{fig:odd_even} Observed odd-even oscillations in O vacancy formation energy ($E_\textrm{Ov}$) as Ti ions are added along the $[110]$, $[001]$ and $[\bar{1}10]$ crystallographic directions of the $(110)$ rutile surface. Starting from structure 1 of panel (a), we created larger clusters by appending additional Ti ions along these directions, whilst keeping the cluster symmetric around the O vacancy -- colored in yellow with a black outline. Calculations were performed at PBE-DFT level with def2-SVP basis set.}
\end{figure}

The poor convergence of the TiO\textsubscript{2} surface with cluster size (Fig.\ 3 (d) in the main text) is a result of the presence of odd-even oscillations in $E_\textrm{Ov}$ as Ti ions are appended along the two crystallographic axes of the TiO\textsubscript{2} surface. These odd-even oscillations in $E_\textrm{Ov}$ are illustrated in Fig.~\ref{fig:odd_even}. Starting from a base quantum cluster (e.g.\ Structure 1), we have appended additional Ti ions along the $[110]$, $[001]$ and $[\bar{1}10]$ directions of the cluster to create larger clusters along those directions, as seen in Fig.~\ref{fig:odd_even} (a). The change in $E_\textrm{Ov}$ for increasing cluster size along these three directions is depicted in Fig.~\ref{fig:odd_even} (b). We observe large oscillation amplitudes in $E_\textrm{Ov}$ in the range of 0.4 and 0.2 eV for the $[\bar{1}10]$ and $[001]$ directions respectively, with no oscillations in the $[110]$ direction.

\FloatBarrier

\newpage
\section{LNO threshold convergence}
Table \ref{tab:lno_thresh} compares the deviation of $E_\textrm{Ov}$ computed at the ``Loose'', ``Normal'' and ``Tight'' LNO threshold settings w.r.t.\ canonical CCSD(T) results for a MgO $(001)$ surface cluster with 5 Mg ions (consisting of the first two RDF shells). Our results indicate that Normal settings give accuracy to within 0.03 eV or less compared to canonical CCSD(T), better than chemical accuracy (0.04 eV). We expect these conclusions to hold for other systems as well and have not performed these tests due to the computational cost of canonical CCSD(T) calculations.

\begin{table}[h]
	\caption{\label{tab:lno_thresh} $E_\textrm{Ov}$ computed at the ``Loose'', ``Normal'' and ``Tight'' LNO threshold settings compared to canonical CCSD(T) results for MgO $(001)$ surface cluster with 5 Mg ions (consisting of the first two RDF shells). Errors are quoted w.r.t.\ canonical results. Calculations were performed using the A'V$X$Z basis set (e.g.\ cc-pV$X$Z and aug-cc-pv$X$Z basis sets on the Mg and O ions respectively) at the double-zeta (DZ) and triple-zeta (TZ) levels. Complete basis set extrapolation with these two basis sets was performed using parameters taken from Neese and Valeev~\cite{neeseRevisitingAtomicNatural2011}.}
	\begin{ruledtabular}
	\begin{tabular}{ldddddd}
		LNO threshold & \multicolumn{1}{c}{\textrm{DZ}}   & \multicolumn{1}{c}{\textrm{Error}} & \multicolumn{1}{c}{\textrm{TZ}}   & \multicolumn{1}{c}{\textrm{Error}} & \multicolumn{1}{c}{\textrm{CBS}} & \multicolumn{1}{c}{\textrm{Error}} \\ \midrule
		Loose         & 6.57 & -0.05 & 6.78 & -0.14 & 6.91       & -0.19 \\
		Normal        & 6.59 & -0.03 & 6.89 & -0.02 & 7.07       & -0.02 \\
		Tight         & 6.60 & -0.02 & 6.91 & -0.01 & 7.09       & 0.00  \\ \midrule
		Canonical     & 6.62 & 0.00  & 6.91 & 0.00  & 7.09       & 0.00 
	\end{tabular}
	\end{ruledtabular}
\end{table}
\FloatBarrier

\section{Final Ov formation energy estimates}

Table \ref{tab:final_eov} shows the final estimates of $E_\textrm{Ov}$ for various DFT levels as well as correlated wave-function methods. Values are given for both the case where the appropriate O\textsubscript{2} binding energy from Table \ref{tab:o2_bind} is used as well as when a correction using the experimental binding energy is applied. As can be seen, the use of the experimental binding energy correction lowers the mean absolute error (MAE) w.r.t.\ LNO-CCSD(T) of the GGA functionals (i.e.\ from 0.85 eV to 0.32 eV for PBE), but this correction does not significantly change the errors for other methods since they match experimental binding energies sufficiently well already (see Table \ref{tab:o2_bind}). The only major exception is LMP2, with errors that increase by $\sim$0.2 eV upon using the experimental binding energy.

\begin{table*}[h]
	\setlength\extrarowheight{-8pt}
	\small
	\caption{\label{tab:final_eov} $E_\textrm{Ov}$ estimates at various levels of DFT XC functional approximations for the SCCs of rocksalt MgO and rutile TiO\textsubscript{2}, in their bulk and common surface planes. The final estimate of the B2PLYP, LMP2, LNO-CCSD and LNO-CCSD(T) are also given. The first half of the table gives the results when the O\textsubscript{2} binding energy corresponding to the appropriate method is used to compute $E_\textrm{Ov}$ whilst the second half gives results when the experimental binding energy of 5.22 eV~\cite{fellerReexaminationAtomizationEnergies1999} is applied instead. Errors are given w.r.t.\ the final LNO-CCSD(T) estimate. The computational details for these calculations can be found in the main text. The final mean absolute error (MAE) row excludes the LMP2 and LNO-CCSD methods.}
	\begin{ruledtabular}
\begin{tabular}{lccccccccc}
\multicolumn{10}{c}{Method binding energy}                                                             \\ \midrule
            & MgO Bulk & Error & MgO Surface & Error & TiO\textsubscript{2} Bulk & Error & TiO\textsubscript{2} Surface & Error & MAE  \\ \midrule
PBE         & 6.66     & -1.02 & 6.15        & -1.03 & 5.67      & -0.72 & 5.00         & -0.55 & 0.83 \\
BP86        & 6.85     & -0.83 & 6.31        & -0.87 & 5.71      & -0.68 & 5.06         & -0.50 & 0.72 \\
BLYP        & 7.11     & -0.56 & 6.54        & -0.64 & 5.66      & -0.73 & 4.91         & -0.64 & 0.64 \\
TPSS        & 6.87     & -0.81 & 6.34        & -0.85 & 5.86      & -0.53 & 5.15         & -0.41 & 0.65 \\
SCAN        & 7.43     & -0.25 & 6.74        & -0.44 & 6.27      & -0.12 & 5.43         & -0.12 & 0.23 \\
PBE0        & 6.99     & -0.69 & 6.39        & -0.79 & 6.07      & -0.32 & 5.41         & -0.14 & 0.49 \\
HSE06       & 7.06     & -0.61 & 6.46        & -0.73 & 6.08      & -0.31 & 5.40         & -0.15 & 0.45 \\
B3LYP       & 7.41     & -0.27 & 6.76        & -0.42 & 6.05      & -0.34 & 5.32         & -0.23 & 0.32 \\
$\omega$B97X& 7.34     & -0.34 & 6.79        & -0.39 & 6.28      & -0.11 & 5.49         & -0.06 & 0.23 \\
B2PLYP      & 7.58     & -0.10 & 7.01        & -0.17 & 6.12      & -0.27 & 5.52         & -0.03 & 0.14 \\
LMP2        & 7.84     & 0.16  & 7.41        & 0.23  & 6.40      & 0.01  & 5.52         & -0.03 & 0.11 \\
LNO-CCSD    & 7.95     & 0.27  & 7.24        & 0.06  & 6.72      & 0.33  & 5.64         & 0.09  & 0.19 \\
LNO-CCSD(T) & 7.68     & --     & 7.18        & --     & 6.39      & --     & 5.55         & --     &      \\
MAE         &          & 0.55  &             & 0.63  &           & 0.41  &              & 0.28  &  \\ \midrule
\multicolumn{10}{c}{Experimental   binding energy}                                                     \\ \midrule
            & MgO Bulk & Error & MgO Surface & Error & TiO\textsubscript{2} Bulk & Error & TiO\textsubscript{2} Surface & Error & MAE  \\ \midrule
PBE         & 7.16 & -0.49 & 6.64 & -0.51 & 6.16 & -0.20 & 5.50 & -0.02 & 0.31 \\
BP86        & 7.31 & -0.34 & 6.77 & -0.38 & 6.18 & -0.19 & 5.52 & 0.00  & 0.23 \\
BLYP        & 7.43 & -0.22 & 6.86 & -0.29 & 5.98 & -0.39 & 5.23 & -0.29 & 0.30 \\
TPSS        & 7.00 & -0.65 & 6.47 & -0.68 & 5.99 & -0.37 & 5.28 & -0.24 & 0.49 \\
SCAN        & 7.57 & -0.08 & 6.89 & -0.26 & 6.42 & 0.06  & 5.58 & 0.06  & 0.11 \\
PBE0        & 7.07 & -0.58 & 6.47 & -0.68 & 6.15 & -0.21 & 5.49 & -0.04 & 0.38 \\
HSE06       & 7.12 & -0.53 & 6.51 & -0.64 & 6.14 & -0.22 & 5.45 & -0.07 & 0.37 \\
B3LYP       & 7.46 & -0.19 & 6.81 & -0.34 & 6.10 & -0.26 & 5.37 & -0.15 & 0.24 \\
$\omega$B97X& 7.43 & -0.22 & 6.88 & -0.27 & 6.37 & 0.01  & 5.58 & 0.06  & 0.14 \\
B2PLYP     & 7.65 & 0.00  & 7.08 & -0.07 & 6.19 & -0.18 & 5.59 & 0.06  & 0.08 \\
LMP2        & 8.06 & 0.41  & 7.63 & 0.48  & 6.62 & 0.26  & 5.74 & 0.21  & 0.34 \\
LNO-CCSD    & 7.74 & 0.09  & 7.03 & -0.12 & 6.51 & 0.14  & 5.43 & -0.09 & 0.11 \\
LNO-CCSD(T) & 7.65 & --    & 7.15 & --    & 6.36 & --    & 5.52 & --    &      \\
MAE         &      & 0.33  &      & 0.41  &      & 0.21  &      & 0.10  &   
\end{tabular}
	\end{ruledtabular}
\end{table*}
\FloatBarrier

\section{Assessing differences to the $\Delta^\textrm{CCSD(T)}_\textrm{PBE}$ method}

\begin{figure}[h]
	\includegraphics{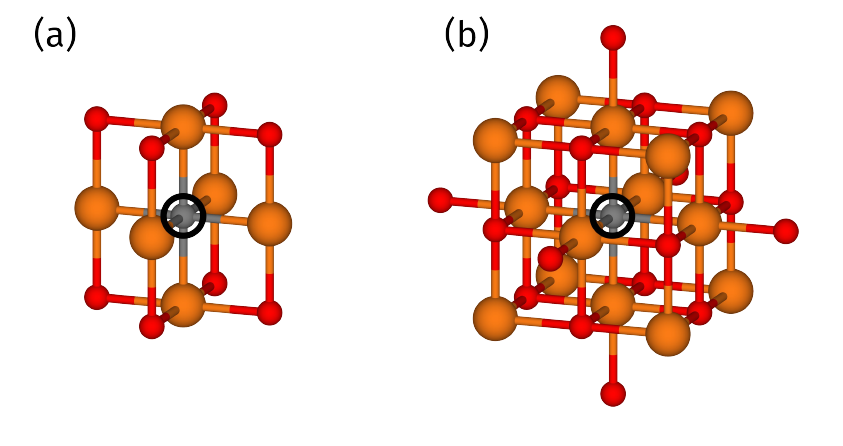}%
	\caption{\label{fig:richtermg} The (a) Mg\textsubscript{6}O\textsubscript{9} and (b) Mg\textsubscript{14}O\textsubscript{19} quantum clusters used to study the Ov in the work of Richter \etal{}~\cite{richterConcentrationVacanciesMetalOxide2013}.}
\end{figure}

In this section, we have performed our own calculations to understand the observed discrepancy of 0.83 eV between our LNO-CCSD(T) $E_\textrm{Ov}$ of 7.68 eV for MgO bulk with the value of 6.85 eV by Richter \etal{}~\cite{richterConcentrationVacanciesMetalOxide2013} using the $\Delta^\textrm{CCSD(T)}_\textrm{PBE}$ approach. The $\Delta^\textrm{CCSD(T)}_\textrm{PBE}$ method of Richter \etal{} determines $E_\textrm{Ov}$ through the embedded cluster approach with the following equation: 
\begin{equation} \label{deltaccsdt}
E_\textrm{Ov}^{\Delta^\textrm{CCSD(T)}_\textrm{PBE}} = E^\textrm{CCSD(T)}_\textrm{Ov} [ \textrm{Mg\textsubscript{6}O\textsubscript{9}}] - E^\textrm{PBE}_\textrm{Ov} [ \textrm{Mg\textsubscript{6}O\textsubscript{9}}] + E^\textrm{PBE}_\textrm{Ov} [ \textrm{Mg\textsubscript{14}O\textsubscript{19}}],
\end{equation}
where the first two terms on the right computes the difference in $E_\textrm{Ov}$ between CCSD(T) and PBE in a small Mg\textsubscript{6}O\textsubscript{9} quantum cluster whilst the last term is the $E_\textrm{Ov}$ for a converged Mg\textsubscript{14}O\textsubscript{19} quantum cluster. Both clusters are visualized in Fig.~\ref{fig:richtermg}. 

In Table~\ref{tab:richtermg}, we have recomputed the $E_\textrm{Ov}$ from the $\Delta^\textrm{CCSD(T)}_\textrm{PBE}$ approach and compared each of the individual terms in Eq.~\ref{deltaccsdt} to assess where differences may arise. The $E^\textrm{PBE}_\textrm{Ov} [ \textrm{Mg\textsubscript{14}O\textsubscript{19}}]$ system used by Richter \etal{} allowed for relaxation of the Mg ions close the Ov, whilst in our work, we have not allowed for this relaxation. If no relaxation around the Ov is used, as we have done in this work, Richter \etal{} would obtain a $E_\textrm{Ov}$ of 6.97 eV compared to 6.85 eV for the relaxed cluster. Thus, relaxation effects in the Mg\textsubscript{14}O\textsubscript{19} cluster accounted for 0.12 eV in the difference w.r.t.\ our LNO-CCSD(T) estimate. Our calculations with LNO-CCSD(T) with ``Tight'' LNO thresholds, using the same A'C'V$X$Z basis set and frozen core treatment as Richter \etal{} produces a $\Delta^\textrm{LNO-CCSD(T)}_\textrm{PBE}$ value of 7.73 eV, which is $\sim 0.1$ eV higher than our best $E_\textrm{Ov}$ estimate using a large 38 Mg ion cluster in Table~\ref{tab:final_eov}. Thus, there is still a large difference w.r.t.\ the values obtained by Richter \etal{} even if the $\Delta^\textrm{CCSD(T)}_\textrm{PBE}$ method is reproduced. 

Considering the individual terms of Eq.~\ref{deltaccsdt} in both our calculation and that of Richter \etal{} in Table~\ref{tab:richtermg}, we find that the the PBE $E_\textrm{Ov}$ value obtained from the Mg\textsubscript{6}O\textsubscript{9} and Mg\textsubscript{14}O\textsubscript{19} clusters are similar between the two studies. There is a increase of 0.08 eV for the two clusters in our study due to the use of different lattice parameters (R2SCAN in this work whilst PBE was used by Richter \etal{}). The main culprit for the different $E_\textrm{Ov}$ values between the two studies lies within the $E_\textrm{Ov}$ value computed with CCSD(T) or LNO-CCSD(T) on the Mg\textsubscript{6}O\textsubscript{9} quantum cluster. 

\begin{table}[h]
	\caption{\label{tab:richtermg} The $\Delta^\textrm{CCSD(T)}_\textrm{PBE}$ computed values of $E_\textrm{Ov}$ from the supplementary material of Richter \etal{} as well as explicit calculations on the same quantum clusters which we performed in this study. }
	\begin{ruledtabular}
	\begin{tabular}{lcccc}
		& $E^\textrm{CCSD(T)}_\textrm{Ov} [ \textrm{Mg\textsubscript{6}O\textsubscript{9}}]$ & $E^\textrm{PBE}_\textrm{Ov} [ \textrm{Mg\textsubscript{6}O\textsubscript{9}}]$& $E^\textrm{PBE}_\textrm{Ov} [ \textrm{Mg\textsubscript{14}O\textsubscript{19}}]$ & $\Delta^\textrm{CCSD(T)}_\textrm{PBE}$ \\ \midrule
		Richter \etal{}~\cite{richterConcentrationVacanciesMetalOxide2013} & 7.09      & 7.18     & 7.06     & 6.97  \\
		This work     & 7.84      & 7.25     & 7.14     & 7.73 
	\end{tabular}
	\end{ruledtabular}
\end{table}

We have shown above that 0.2 out of the 0.8 eV difference between our best LNO-CCSD(T) estimate and the $\Delta^\textrm{CCSD(T)}_\textrm{PBE}$ estimate of Richter \etal{} arises due to differences in lattice parameter and inclusion of relaxation effects by Richter \etal{}. The remaining 0.6 eV gap could arise from differing electrostatic embedding environments (e.g.\ point charge and ECP regions) between the two studies. Our work and others~\cite{chenColorCenterSinglet2020} have indicated that cWFT methods are highly sensitive to the electrostatic embedding environment compared to DFT and this requires careful convergence, such as those in Table \ref{mgo_bulk_size} or Fig.\ 4 of the main text.